\documentclass[%
 aip,
 rsi,
amsmath,amssymb,
reprint,
floatfix
]{revtex4-2}
\usepackage{graphicx}% Include figure files
\usepackage{dcolumn}% Align table columns on decimal point
\usepackage{bm}% bold math
\usepackage[utf8]{inputenc}
\usepackage{mathptmx}
\usepackage{multirow}
\usepackage{braket}
\usepackage[english]{babel}
\usepackage{xcolor}% Include colors for document elements
\colorlet{RED}{red}
\colorlet{BLUE}{blue}
\usepackage{dcolumn}% Align table columns on decimal pointhttps://v1.overleaf.com/17481932tmwjwzytcbcq#
\usepackage{bm}% bold math
\usepackage[version=4]{mhchem} % HvD: for general chemical formula like \ce{(TiO2)n}
\usepackage{acronym}
\usepackage{adjustbox}
\usepackage{float}
\usepackage{tikz}
\usepackage{microtype} % fixes many of the overflow hbox errors
\usetikzlibrary{calc,shapes.geometric,decorations.pathmorphing,patterns}

\definecolor{background-color}{gray}{0.98}
\usepackage[margin=2.3cm,bmargin=1cm,footnotesep=1cm]{geometry}

\begin{document}

\title{
%Searching a compact unitary basis for the quantum simulation of non-unitary theories
Mapping renormalized coupled cluster methods to quantum computers through a compact unitary representation of non-unitary operators
}

\author{Bo Peng}
\email{peng398@pnnl.gov}
 \affiliation{Physical Sciences and Computational Division, Pacific Northwest National Laboratory, Richland, Washington 99354, United States of America}
\author{Karol Kowalski} 
\email{karol.kowalski@pnnl.gov}
 \affiliation{Physical Sciences and Computational Division, Pacific Northwest National Laboratory, Richland, Washington 99354, United States of America}

\begin{abstract}
Non-unitary theories are commonly seen in the classical simulations of quantum systems. 
Among these theories, the method of moments of coupled-cluster equations (MMCCs) and 
the ensuing classes of the renormalized coupled-cluster (CC) approaches have evolved into one of 
the most accurate approaches to describe correlation effects in various quantum systems. 
The MMCC formalism provides an effective way for correcting energies of approximate CC formulations 
(parent theories)  using moments, or CC equations, that are not used to determine approximate cluster 
amplitudes. In this paper, we propose a quantum algorithm for computing MMCC ground-state energies that  
provide two main advantages over classical computing or other quantum algorithms: (i) the possibility of 
forming superpositions of CC moments of arbitrary ranks in the entire Hilbert space and using an arbitrary 
form of the parent cluster operator for MMCC expansion; and (ii) significant  reduction in the number of 
measurements in quantum simulation through a compact unitary representation for a generally 
non-unitary operator. We illustrate the robustness of our approach over a broad class of test cases, 
including  $\sim$40 molecular systems with varying basis sets encoded using 4$\sim$40 qubits, 
and exhibit the detailed MMCC analysis for the 8-qubit half-filled, four-site, single impurity Anderson model 
and 12-qubit hydrogen fluoride molecular system from the corresponding noise-free and noisy MMCC quantum computations. 
We also outline the extension of MMCC formalism to the case of unitary CC wave function ansatz.
\end{abstract}

\maketitle

\section{Introduction}
\label{section1}

Enabling accurate calculations for strongly correlated quantum systems using many-body methodologies 
that can adequately capture collective many-body effects is the focus of ongoing efforts in nuclear theory, 
condensed matter physics, materials sciences, and quantum chemistry. To this end, over the last decades 
several promising formulations, including density functional theory, Green's function method, and 
wave function approaches have been intensively developed and tested in the context of enabling computational 
tools to deal with complex processes involving strong correlation effects. Due to the possibility of 
constructing hierarchical classes of approximations approaching the exact (or full configuration interaction [FCI]) limit, 
the wave-function-based methods are especially attractive and have evolved into 
the most accurate formulations currently used in many-body physics and chemistry. 
For this reason,  the coupled cluster (CC) formalism,
\cite{coester58_421,coester60_477,cizek66_4256,paldus72_50,mukherjee1975correlation,purvis82_1910,bishop1987coupled,paldus07,crawford2000introduction,bartlett_rmp}
due to its ability to ensure a proper scaling of ground-state energy with the number of particles, 
has assumed special position in high-accuracy studies of quantum systems.

The inception of the CC theory 
has instigated  numerous efforts toward
the development of approximations capable of providing a high level of accuracy for calculated energies. 
These formulations played an immense role in understanding 
fundamental problems in many-body physics
 \cite{arponen83_311,arponen1987extended1,arponen1987extended,arponen1991independent,arponen1993independent,arponen1993independent,
robinson1989extended,arponen1991holomorphic,bishop1978electron,bishop1982electron,arponen1988extended,emrich1984electron} 
(for an excellent review of these developments, see Ref.  \citenum{bishop1991overview}),
quantum field theory,\cite{funke1987approaching,kummel2001post,hasberg1986coupled,bishop2006towards,ligterink1998coupled} quantum hydrodynamics,\cite{arponen1988towards,bishop1989quantum}
nuclear structure theory,\cite{PhysRevC.69.054320,PhysRevLett.92.132501,PhysRevLett.101.092502,hagen2014coupled} quantum chemistry,\cite{scheiner1987analytic,sinnokrot2002estimates,slipchenko2002singlet,tajti2004heat,crawford2006ab,parkhill2009perfect,riplinger2013efficient,yuwono2020quantum,PhysRevX.10.041043} 
and condensed phase systems.
\cite{stoll1992correlation,hirata2004coupled,katagiri2005equation,harl2009accurate,booth2013towards,zhang2019coupled,
degroote2016polynomial,mcclain2017gaussian,wang2020excitons,
farnell2004coupled,farnell2018interplay,bishop2019frustrated}
For single-reference (SR) CC formulations, where one can approximate the exact wave function $|\Psi\rangle$ 
by a single determinant $|\Phi\rangle$, the CC ansatz given by the expansion 
\begin{equation}
    |\Psi\rangle = e^T |\Phi\rangle
    \label{in1}
\end{equation}
can be used to define a hierarchy of approximations converging to the FCI results, where the cluster operator $T$
\textcolor{black}{is defined as the sum of its many-body components $T_k$
\begin{equation}
T=\sum_{i=1}^{N} T_k \;,
\label{cc2}
\end{equation}
with $T_k$ producing $k$-tuple excitations  when acting on the reference function $|\Phi\rangle$. 
The second quantized from $T_k$ is defined as
\begin{equation}
T_k = \frac{1}{(k!)^2} \sum_{i_1,\ldots,i_k; a_1,\ldots, a_k} t^{i_1\ldots i_k}_{a_1\ldots a_k} 
a^{\dagger}_{a_1} \ldots a^{\dagger}_{a_k} a_{i_k} \ldots a_{i_1}\;
\label{T_k}
\end{equation}
with $a_p^{\dagger}$ ($a_p$) the creation (annihilation) operators for electron in $p$-th state and indices $i_1,i_2,\ldots$ ($a_1,a_2,\ldots$) referring to occupied (unoccupied) spin-orbitals in the reference function. }
The cluster operator $T$ can be truncated at some excitation level with respect to the reference Slater determinant $|\Phi\rangle$.  
For example, the ubiquitous 
\textcolor{black}{CC approach with single and double excitations (CCSD),\cite{purvis82_1910} CC approach with single, double, and triple excitations (CCSDT),\cite{ccsdt_noga,ccsdt_noga_err,scuseria_ccsdt}
and CC approach with single, double, triple, and quadruple excitations (CCSDTQ)  \cite{Kucharski1991,ccsdtq_nevin} methods}
are obtained by truncating the cluster operator expansion at double, triple, and quadruple excitation levels, respectively. 
%In this way, one can define a hierarchy of methods where one include cluster operators up to certain rank of excitation, 
%i.e., singles (represented by single excitation $T_1$ with respect to the reference function $|\Phi\rangle$), doubles ($T_2$), 
%triples ($T_3$), quadruples ($T_4$), etc. For example, the CCSD method is defin $T\simeq T_1+T_2$,  CCSDT  by $T\simeq T_1+T_2+T_3$,
%  CCSDTQ by $T\simeq T_1+T_2+T_3+T_4$, etc. 
These methods, along with their perturbative variants for economic inclusion of higher rank clusters driven by 
linked-cluster theorem, \cite{brueckner1955many,goldstone1957derivation,brandow67_771} form a basis for a highly accurate 
description of correlation effects in quantum chemistry and nuclear physics. Methods like CCSD(T),\cite{raghavachari89_479} 
which combines iterative CCSD methods with linked cluster theorem to determine triply excited cluster operators,  
and their  implementations are now used in numerical simulations capable of taking advantage of leadership-class 
parallel architectures to describe ground-state correlation effects in a broad class of quantum many-body systems in chemistry 
and condensed phase physics. The $e^T$ operator in Eq. (\ref{in1}) is often referred to as the wave operator,  
which in the SR-CC case is non-unitary.

While the SR-CC is a method of choice for closed-shell systems, for systems characterized by a multi-configurational wave function, due to problems with capturing static correlation effects,  these approaches fail to reproduce the desired accuracy level in calculated ground-state energies.
%due to the inability of a single Slater to describe  a wide spectrum of complicated static correlation effects. 
These effects can be potentially captured by properly defined multi-reference (MR) variants of CC theory, \cite{mrcclyakh} 
which extend the notion of reference function $|\Phi\rangle$ to the model space $\mathcal{M}_0$ to provide a zeroth-order 
description of the wave function. This resulted in several classes of MR-CC theories in Hilbert \cite{jezmonk,mrcclyakh} 
and Fock \cite{mukherjee1975correlation,pal1988molecular,jeziorski1989valence,kaldor1991fock,bernholdt1999critical,mrcclyakh} 
spaces and in the form of canonical transformation theory, \cite{canonical1} where energies are furnished by diagonalization 
of the effective Hamiltonians in model spaces. These developments are also an outcome of earlier intensive attempts toward 
defining efficient MR methodologies that are critically needed to describe strongly correlated systems. 
Over the last few  decades, a significant effort 
\cite{mukherjee1986linked,ims1,meissner1995effective,meissner1998fock,landau1999intermediate,gms1,mahapatra1,hanauer2011pilot,kohn2013state,evangelista2011orbital,evangelista2012sequential,canonical1,canonical2,datta2011state,demel2013additional} 
has been exerted to alleviate challenges that are commonly attributed to the so-called intruder-state problem,\cite{schucan1,schucan2}
which limits the applicability of the MR-CC theories.

An important formulation of  CC theory that applies both to SR and MR cases is provided by 
the methods of moments of coupled-cluster equations (MMCCs). \cite{bookmmcc,mmcc1,mmcc2,kowalski2004new,pittner2009method}  
This formalism allows one to construct an asymmetric energy functional, which can be viewed as a measure of error committed in the approximate SR- and MR-CC calculations and provides a theoretical platform to construct non-iterative energy corrections using non-vanishing moments of CC formulations. It has been demonstrated that MMCC expansions can remove divergences of many-body perturbation theory due to inadequate choice of the reference function (SR case) or the presence of the intruder state problem (MR case). 
The efficacy of this formalism is also tied to the quality of the so-called trial wave functions that contain excitations corresponding to the non-vanishing moments of the approximate CC formalism.
The MMCC methods and ensuing CC renormalized methodologies provide means to control the accuracy of corresponding energies 
by including higher rank moments and/or using more accurate sources of the trial wave functions. 
The completely renormalized CC methods usually build on the approximate theories such as the CCSD and CCSDT formulations. 
For these situations, the moments are homogeneous and do not require convoluted logic, which is sometimes 
the case when adaptive (or configuration selective) representations of the trial wave functions are employed. 
For example, correcting CCSD energies usually requires triply and/or quadruply excited moments ($M_3$/$M_4$), 
while for the CCSD(T) case, $M_4$ is typically used. 
However, the inclusion of  all possible moments to take advantage of full-length moment expansion
%for any type of approximate formalism 
poses a considerable numerical challenge. The renormalized CC methods also offer a unique chance to combine 
the most appealing features of the configuration interaction and CC theories. This has been corroborated in a series of studies, 
including the most recent development employing Quantum Monte-Carlo FCI formalism.
\cite{deustua2017converging,deustua2018communication,deustua2019accurate}
The results obtained with this combined framework are among the most accurate approximate approaches currently used in molecular and nuclear structure calculations. 

%A common feature of all CC methods mentioned earlier is their steep 

\textcolor{black}{Nevertheless,} the numerical overhead associated with calculating CC moments in the canonical representation 
is a steep function of the number of correlated particles and the number of so-called basis set functions 
used to discretize the problem (both  factors are customarily identified as a ``system size" - $N$), 
which prohibits a practical utilization of the formulations based on the high-rank moments. 
For example, for the CCSD parent formalism, the cost of obtaining triply ($M_3$), quadruply 
($M_4$), pentuply ($M_5$), and hextuply ($M_6$) excited moments using classical computing algorithms are proportional to $N^7$, $N^9$,
$N^{11}$, $N^{13}$, respectively. For the CCSDT parent approach, the cost of calculating non-vanishing moments is even higher. 
This scaling precludes the full utilization of  trial wave functions
(for example, very accurate  Quantum Monte-Carlo FCI wave functions \cite{booth2009fermion,booth2013towards}) 
that carry the information about excitations beyond the rank of numerically computable moments. 

Novel computational paradigms associated with the emergence of quantum computing has the potential 
to significantly extend the applicability of variants of the CC formalism, especially unitary.
\cite{pal1984use,hoffmann1988unitary,unitary1,unitary2,kutzelnigg1991error,sur2008relativistic,cooper2010benchmark,Evangelista2019Exact,izmaylov2020order,anand2022quantum}
%{\color{red} more here}
%especially their unitary CC variants \cite{...} in the context of the hybrid Variational Quantum Eigensolvers (VQE) \cite{...}.
%
\textcolor{black}{Typically, there are two distinct quantum approaches, variational Quantum Eigensolver (VQE) methods
\cite{peruzzo2014variational,mcclean2016theory,romero2018strategies,PhysRevA.95.020501,Kandala2017,kandala2018extending,PhysRevX.8.011021,huggins2020non,Cao2019Quantum,Grimsley2019adaptive,grimsley2019trotterized,verteletskyi2020measurement,mcardle2019variational,mcardle2020quantum,tilly2021variational} 
and Quantum Phase Estimator (QPE) formalisms,
\cite{Seeley2012Bravyi,Haner2016High,Wecker2015Progress,Poulin2018Quantum,Child2010Relationship,Cleve1998Quantum,Berry2007Efficient,Luis1996Optimum} 
which use various types of quantum information manipulations for different phases of quantum hardware development. 
While the hybrid VQE formulations target near-term noisy intermediate-scale quantum (NISQ) devices characterized 
by long enough coherence times to perform specific operations on quantum data, the QPE methods, 
which represent qubit encoding of the unitary time evolution of quantum systems, 
are slated for more mature error-corrected quantum devices where circuit-depth limitations can be effectively handled. 
A variety of theoretical methods have been developed to enable VQE and QPE simulations for realistic many-body systems, 
including formulations ranging from Trotterization techniques, general error control formulations for 
Trotter product expansion for real- and imaginary-time dynamics,\cite{Childs2021Theory} 
qubitization/block-encoding methods, \cite{Low2019hamiltonian,Childs2012Quantum} 
truncated Taylor series for simulating Hamiltonian dynamics,\cite{Berry2015Simulating} 
adaptive unitary expansions,\cite{Grimsley2019adaptive} and disentangled form of unitary CC expansions.\cite{Evangelista2019Exact} 
These advances paved the way for early quantum algorithms for simulating quantum dynamics, many-body Green’s function, 
quantum embedding procedures, ground- and excited-state methodologies, and optimal utilization of orbital basis 
(for a detailed review of applications, see Ref. \citenum{Cao2019Quantum}). In contrast to the QPE formalism, 
the VQE method can also render wave function parameters, which makes VQE an easy-to-integrate component of 
multi-scale/multi-physics embedding workflows.\cite{Ma2020Quantum} The availability of circuit-optimized variants of 
the VQE and QPE formalisms will further enhance the ability of the scientific community to 
generate results unobtainable with classical computers. Nevertheless, the VQE optimization is NP-hard, 
and sometimes not robust for processing many wave function parameters to capture static and dynamical correlations effects.}
%For example, Variational Quantum Eigensolvers  (VQE) techniques  are capable of taking advantage of Noisy Intermediate-Scale Quantum (NISQ) devices in simulations of chemical and physical systems. 
An important step toward addressing practical application of the VQE was development of the qubit CC ansatzes, 
\cite{ryabinkin2018qubit} where instead of explicit Fermionic construction of the unitary coupled cluster (UCC) ansatz, 
one builds directly in the qubit space. 

We believe that analogous ideas may be employed to implement quantum algorithms for a broad class of MMCC methods %
\textcolor{black}{based on the flexible forms (either unitary or non-unitary) of the wave operator. In this direction,
one of the major hurdles is to encode the non-unitary operator on quantum computers that only accept unitary forms.
Typically, an arbitrary non-unitary operator (mainly the Fermionic operator in the present discussion) 
can be transformed into a linear combination of Pauli terms. Since each Pauli term is a unitary operation, 
this is the basic version of a linear combination of unitaries. If considering a fault-tolerant implementation 
with favorable resource scaling, a general non-unitary operator can be encoded through a linear combination of 
unitaries technique,\cite{Childs2012} amplitude amplification approach,\cite{brassard2002quantum}  
Hamiltonian simulation,\cite{Berry2015Simulating} qubitization,\cite{low2017optimal} or the direct 
block-encoding\cite{gilyen2019quantum} methods at the cost of introducing deeper circuit and implementing 
controlled-$\hat{\mathcal{U}}$ operations, which nevertheless come with a probability of failure and 
require advanced circuit and error mitigation that might go beyond the capability of the current NISQ devices.} 

\textcolor{black}{Toward a more feasible NISQ approach, Izmaylov et al. proposed a unitary partitioning scheme \cite{Izmaylov2019} 
using numerical graph analytic tools to pre-process a linear combination of Pauli terms for their equivalent encoding on a quantum circuit. 
In light of this approach, here
we propose a more efficient unitary partitioning approach guided by the SR used in the simulation.
In particular, we found that the non-unitary wave operator when acting 
on a SR trial wave function can be represented by a much more compact unitary basis, 
thus provides a more efficient route for performing the general non-unitary quantum simulations.
Based on this observation, we further propose
a quantum algorithm for measuring renormalized CC energies corresponding to 
approximate CC formulations with a non-unitary wave operator on quantum computers. 
Our quantum algorithm exhibits two main advantages over classical computing or other quantum algorithms: 
(i) it provides a framework for introducing new classes of MMCC formulations that includes all possible CC moments
corresponding to flexible forms of the trial wave function; and (ii) the compact unitary representation of a 
general non-unitary operator significantly reduces the number of measurements in quantum simulation.}

%Our algorithm draws heavily on 
%a new therapy combining with graph theory searching for 
%using a compact unitary basis for a general non-unitary operator to deal with the non-Hermitian character of 
%the MMCC formalism and reduces the number of measurements. 
%The new strategy maps the unitary partitioning problem to a graph ``minimum clique cover" problem, builds a graph based on the Pauli terms and their corresponding anti-commuting relationship, and utilizes the graph tools for the unitary search. 
%Numerical tests have been conducted for a wide range of systems to show the capability of 
%exponentially compressing the conventional Pauli terms to 
%representing not only the Hamiltonian, but more generally and importantly, the non-unitary wave operator when acting 
%on a SR trial wave function, in a much more compact unitary basis, 
%thus providing an efficient route for performing the general non-unitary quantum simulations.
%As will be demonstrated, the advantage of using quantum computing  lies in the fact that moments of 
%arbitrary rank can be efficiently mapped to register qubits.
%Since the discussed quantum algorithm can be  used in the context of general wave operator formalism, 
%we also extend the discussion to the MMCC formalism \textcolor{black}{employing} the unitary CC ansatz.
%
\textcolor{black}{This paper is organized as follows. Section \ref{section2} will briefly review the 
MMCC formulation. Section \ref{section3} will detail our proposed quantum algorithm for computing the MMCC energy. 
In Section \ref{section4}, we will first briefly review the original unitary partitioning approach, 
and test its numerical performance over a broad range of molecular systems. We then detail our proposed unitary partitioning approach 
and give some preliminary comparison with respect to the original approach. More comprehensive numerical demonstrations of
our unitary partitioning approach and the associated quantum algorithm are given in Section \ref{section5}, 
where the quantum simulations are conducted for a wide range of molecular systems in both noise-free and noisy environments. 
We will conclude this paper in Section \ref{section6}, where some future developments will also be briefly discussed. }

%%%%%%%%%%%%%%%%%%%%%%%%%%%%%%%%%%%%%%%%%%%%%%%%%%%%%%%%%%%%

\section{Many-body Formulation of the Method of Moments of Coupled Cluster Equations}
\label{section2}

The MMCC equations and ensuing class of renormalized CC formalisms
\cite{bookmmcc,mmcc1,mmcc2,kowalski2005extensive,creom,crcc,cu2o2,piecuch2006single,crccbb,crccopen,crccobb,crccrev,kowalski2018regularized,deustua2017converging,deustua2018communication,deustua2019accurate,eriksen2020ground,gururangan2021high}
have evolved into one of the most accurate methodologies for the high-precision evaluation of ground-state energies for chemical and nuclear systems. 
The idea of the MMCC approach is based on the asymmetric energy functional
\begin{equation}
    E_{\rm MMCC}[\Psi_T]=\frac{\langle\Psi_T|He^{T^{(A)}}|\Phi\rangle}
    {\langle\Psi_T|e^{T^{(A)}}|\Phi\rangle}
    \label{eq1}
\end{equation}
where $|\Psi_T\rangle$ is the so-called trial wave function,  $T^{(A)}$ is an arbitrary approximation (the parent approach) 
to the exact cluster operator $T$, and $H$ is a many-body  Hamiltonian defined by one- and two-body interactions. 
\textcolor{black}{The word ``asymmetric" here refers to the the trial wave function $\langle\Psi_T|$ and 
the wave function $^{T^{(A)}}|\Phi\rangle$ are not complex conjugates of each other.} When $|\Psi_T\rangle$  
is replaced by the exact ground-state wave function $|\Psi\rangle$, 
then the value of the MMCC functional obtained from Eq. (\ref{eq1}) is equal to the exact ground-state energy $E$;
\begin{equation}
    E_{\rm MMCC}[\Psi]=E \;.
    \label{eq2}
\end{equation}
The above functional can be rewritten in a  moment-explicit form by introducing resolution of identity 
$e^{T^{(A)}}e^{-T^{(A)}}$ in the  numerator of MMCC functional
\begin{eqnarray}
 E_{\rm MMCC}[\Psi_T] &=& \frac{\langle\Psi_T|He^{T^{(A)}}|\Phi\rangle}
    {\langle\Psi_T|e^{T^{(A)}}|\Phi\rangle} \notag \\
    &=& \frac{\langle\Psi_T|e^{T^{(A)}}e^{-T^{(A)}}He^{T^{(A)}}|\Phi\rangle}
    {\langle\Psi_T|e^{T^{(A)}}|\Phi\rangle} \notag \\
    &=& \frac{\langle\Psi_T|e^{T^{(A)}}M^{(A)}|\Phi\rangle}
    {\langle\Psi_T|e^{T^{(A)}}|\Phi\rangle} \;,
    \label{eq3}
\end{eqnarray}
where the action of  moment operator $M^{(A)}$ on the reference function is defined as 
\begin{equation}
   M^{(A)}|\Phi\rangle=  e^{-T^{(A)}}He^{T^{(A)}}|\Phi\rangle  \;.
    \label{eq4}
\end{equation}
The many-body form of the above functional  for the trial wave function, representing the exact one,  
allows for finding the relationship between  exact energy $E$ and the energy $E^{(A)}$ of the approximate 
CC formulation defined by an arbitrary $T^{(A)}$ cluster operator. 
To show this, let us decompose the unit operator $I$ acting in the Hilbert space into 
$P$, $Q_A$, and $Q_R$ parts, 
\begin{equation}
    I=P+Q_A+Q_R
    \label{eq5}
\end{equation}
where $P=|\Phi\rangle\langle\Phi|$ is a projection operator onto the reference function $|\Phi\rangle$, 
$Q_A$ is a projection operator  onto a sub-space of excited configurations (with respect to the reference function) 
generated by action of $T^{(A)}$ onto the reference function, and $Q_R$ is a projection operator 
onto all remaining excited Slater determinants. Next, let us expand the numerator of Eq. (\ref{eq3})
(with $|\Psi_T\rangle$ replaced by $|\Psi\rangle$)
\begin{eqnarray}
    E &=& \frac{\langle\Psi|e^{T^{(A)}}M^{(A)}|\Phi\rangle}
    {\langle\Psi|e^{T^{(A)}}|\Phi\rangle} \notag \\
    &=& \frac{\langle\Psi|e^{T^{(A)}}(P+Q_{A}+Q_{R})M^{(A)}|\Phi\rangle}
    {\langle\Psi|e^{T^{(A)}}|\Phi\rangle} \notag \\
    &=& E^{(A)}+\frac{\langle\Psi|e^{T^{(A)}} Q_R M^{(A)}|\Phi\rangle}
    {\langle\Psi|e^{T^{(A)}}|\Phi\rangle} \;,
    \label{eq6}
\end{eqnarray}
where the last term on the right-hand side of Eq. (\ref{eq6}) gives the algebraic form of the correction 
that needs to be added to $E^{(A)}$ to recover the exact energy $E$. For this reason, this error formula 
was used to define approximate non-iterative forms  of  corrections to improve the quality of the $E^{(A)}$ energies. 
For non-trivial forms of the  corrections, the trial wave functions  have to incorporate excitations from the $Q_R$ sub-space. 
In the derivation of Eq. (\ref{eq6}), we also assumed that the cluster amplitudes defining $T^{(A)}$ are obtained by 
solving CC equations
\begin{equation}
    Q_A M^{(A)}|\Phi\rangle = 0 \;. 
    \label{eq7}
\end{equation}
Other scenarios, where one uses other sources for evaluating  $T^{(A)}$ amplitudes, are also possible. In such  situations, all 
$Q_A M^{(A)}|\Phi\rangle$ terms have to be included in the numerator of Eq. (\ref{eq6}). 

In the last two decades, several variants of the MMCC  or renormalized CC methods involving various sources of the trial wave functions and different rank moments have been tested in challenging situations corresponding to bond-breaking and bond-forming processes 
where standard perturbative approaches such as CCSD(T) or CCSD(TQ) methods tend to provide non-physical shapes of ground-state potential energy surfaces (PESs). 
It was demonstrated that the renormalized CC methods could significantly improve the quality of these methods, 
especially for quasidegenerate electronic states. One of the most efficient formulations of 
renormalized CC methods uses a special form of the trial wave function originating in the 
bi-variational character of the CC theory,
 \cite{arponen1987extended,Piecuch2005Renormalization} i.e., 
\begin{equation}
    \langle\Psi_T| = \langle\Phi| \mathcal{L}_T e^{-T^{(A)}} \;,
    \label{eq7a}
\end{equation}
where  $\mathcal{L}_T$ is the trial left de-excitation operator
\textcolor{black}{defined as
\begin{align}
\mathcal{L}_T = 1 + \sum_{n=1}^N \Lambda_{T,k}. \label{L_T}
\end{align}
Here
\begin{align}
\Lambda_{T,k} = \frac{1}{(k!)^2} \sum_{i_1,\ldots,i_k; a_1,\ldots, a_k} l_{i_1\ldots i_k}^{a_1\ldots a_k} 
a^{\dagger}_{i_1} \ldots a^{\dagger}_{i_k} a_{a_k} \ldots a_{a_1}\; \label{L_Tn}
\end{align}
are the $k$-body components of $\mathcal{L}_T$. Eqs. (\ref{L_T}) and (\ref{L_Tn}) 
}
%(which is a consequence of bi-variational \cite{arponen1987extended}
%character of the CC formalism), 
lead to a very simple, denominator-free form of the asymmetric energy  functional
\begin{equation}
    E[\mathcal{L}_T]=\langle\Phi|\mathcal{L}_T M^{(A)}|\Phi\rangle \;.
    \label{eq8}
\end{equation}
This class of renormalized CC methods (termed CR-CC($m,n$)\cite{piecuch2006single,shenpp3} approximations) 
along with the older MMCC formulations based on the numerator-denominator expansion \cite{kowalski2005extensive} 
are size-consistent, thus providing proper (additive) separability of the energies of composite systems 
in the non-interacting limit. This fact plays a key role in describing chemical reactions and in proper 
descriptions of systems in the thermodynamic limit.
%\textcolor{black}{The MMCC formation can also be extended to employ unitary anstaz }

\section{MMCC Quantum Algorithm}
\label{section3}

\textcolor{black}{Note that the MMCC formulations employing either unitary or non-unitary CC ansatze can be evaluated using 
classical computers by treating tensor contractions, while the unitary is best for quantum computation.}
In this section, we outline the main features of the quantum algorithm for 
calculating standard moment corrections (\ref{eq1}), \textcolor{black}{in particular, treating the non-unitary cases}.
We will entirely focus on the standard MMCC formulations to emphasize critical steps defining the algorithms. 
The extensions to alternative ansatzes (for example, UCC ansatz) can be achieved 
in a similar way \textcolor{black}{(see Appendix \ref{MMUCC})}. 
In contrast to the algorithms used for calculating moments on classical computers, 
we will assume a different strategy. Instead of calculating the connected form of the moments 
given by Eq. (\ref{eq4}) directly, we will represent the following vectors 
(needed to calculate the MMCC correction) as a sum of Pauli strings, $P_s$, 
or general unitaries, $U_l$, (see the following section for technical details):
\begin{eqnarray}
e^{T^{(A)}}|\Phi\rangle & \rightarrow & |\Omega\rangle =\left( \sum_l \omega_l U_l \right) |\Phi\rangle  \;,\label{q3} \\
He^{T^{(A)}}|\Phi\rangle &\rightarrow& |\Gamma\rangle = \left( \sum_l
\gamma_l U_l  \right)|\Phi\rangle \;, \label{q1} \\
|\Psi_T\rangle &\rightarrow& |\Theta\rangle =\left( \sum_l \theta_l U_l  \right) |\Phi\rangle\label{q2} 
\end{eqnarray}
where $H=\sum_l h_l U_l $ and coefficients $\lbrace \gamma_l \rbrace$ depend on  coefficients $\lbrace \omega_l \rbrace $ and  $\lbrace  h_l \rbrace$. Calculating MMCC requires measuring two overlaps 
\begin{equation}
    \langle\Theta|\Gamma\rangle \;\;\;{\rm and} \;\;\;
    \langle\Theta|\Omega\rangle
    \label{q3}
\end{equation}
which can be obtained by using the Hadamard-type measurement for expectation values of 
the products of unitaries (which are also unitaries). 
%This representation involves a  direct evaluation of target Ansatz and trial wave functions, which leads to their explicit forms compared to the ``classical'' strategy where moments are evaluated first. 
There are several advantages of this algorithm:
\begin{itemize}
    \item The non-normalized state $|\Gamma\rangle$ contains superposition of moments of all ranks,
%    (in term of the rank  excitation) 
    defining the expression  
    $e^{T^{(A)}}M^{(A)}|\Phi\rangle$
    (see second line in Eq. (\ref{eq3})). 
    In the typical implementation of renormalized CC methods, one deals with low-rank methods, 
    for example triply and quadruply excited moments for $T^{(A)}$ defining the CCSD approach. 
    In this sense, our proposed quantum algorithms enable introduction of new classes of  MMCC formulations  
    that include all possible CC moments. 
    \item The discussed representation can be used to calculate moments corresponding to any form of  $T^{(A)}$, 
    including some adaptive CC parametrizations 
    (for example, the adaptive CC expansions of Ref. \citenum{lyakh2010adaptive}),
    which poses a significant challenge for their efficient implementations on classical computers.
    \item The above algorithm is well-posed to use the wave function representations for  trial wave function(s) 
    that cannot be easily implemented on classical machines. This includes various UCC  and 
    imaginary-time representations of the wave function (e.g., $ N(\tau) e^{-\tau H}|\Phi\rangle$ 
    with $N(\tau)$ being the normalization factor) recently explored in the context of quantum algorithms in 
    Refs. \citenum{mcardle2019variational} and \citenum{motta2020determining}. The later representation is also closely 
    related to the recent implementations of the MMCC formalisms using Quantum Monte-Carlo FCI formalism representation of 
    the trial wave function 
    \cite{booth2009fermion}
    and the truncated-rank moments expansion.\cite{deustua2017converging}
    \item The two-step character of the MMCC formulations, composed of 
    (i) calculating the cluster amplitudes for the parent formulation defined by the $T^{(A)}$ operator and 
    (ii) forming the MMCC correction, is well suited to guide the process of eliminating some errors in step (ii) by error analysis in cheaper step (i). 
\end{itemize}
Also, note that if $\langle\Theta|$ corresponds to $\langle\Phi|\mathcal{L} e^{-T^{(A)}}$,
then only $\langle\Theta|\Gamma\rangle$ overlap needs to be calculated. 

The efficiency of the quantum algorithm for MMCC formalism critically depends on our ability to approximate the algebraic form of $|\Omega\rangle$, $|\Gamma\rangle$, and $|\Theta\rangle$ vectors and compress their qubit representations. Therefore, 
in the following sections, we will provide basic tenets of the implementation that requires 
encoding non-unitary and non-Hermitian operators. 

In general, the MMCC theory and asymmetric energy functionals  can be universally applied to \textcolor{black}{\textit{any}} form of the wave operator, 
\textcolor{black}{either unitary or non-unitary, }
acting on any reference function, which can be represented as a superposition of several Slater determinants. 
In some cases, where one deals with a truly \textcolor{black}{non-unitary} entangled reference state, 
using our proposed quantum algorithm can be particularly beneficial.  
%
% the structure of the MMCC formalism provides also a natural mechanism for error mitigation techniques in quantum simulations of the MMCC energy corrections. 
%

%
% similar form of algorithms for unitary cCC
%

\section{Representation of non-Unitary operators on quantum computers}\label{section4}

\begin{figure*}
    \centering
    \includegraphics[width=\linewidth]{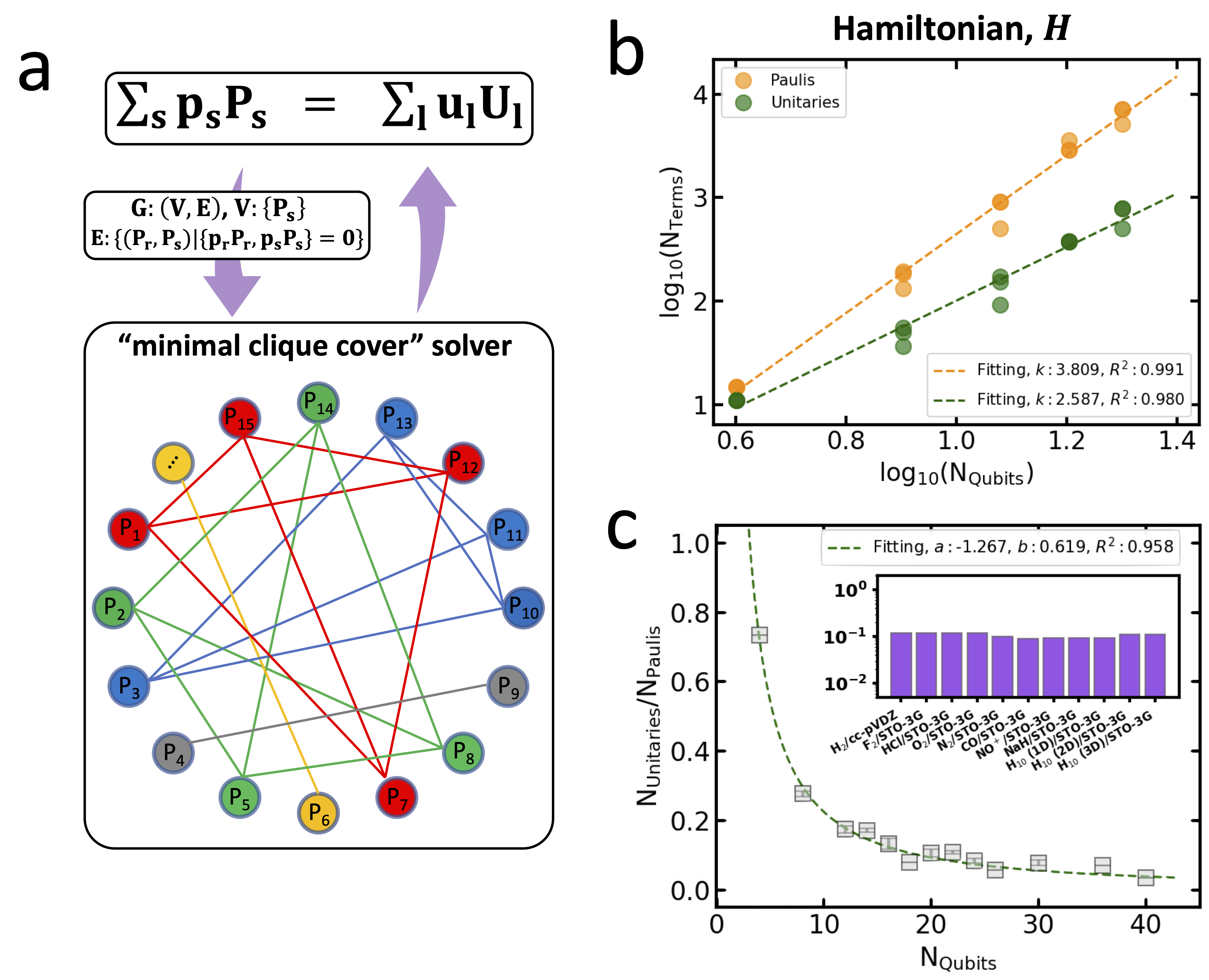}
    \caption{(a) Schematic workflow of the unitary partitioning approach proposed in Ref.  \citenum{Izmaylov2019}. 
    (b) Numbers of Pauli terms ($N_{\rm Paulis}$) and corresponding numbers of unitary groups ($N_{\rm Unitaries}$) 
    determined from the proposed anti-commuting rule for the Hamiltonians of 1D, 2D, and 3D hydrogen systems encoded 
    using up to 20 qubits. The geometries of these hydrogen systems are given in supplementary materials, 
    and the nearest H$-$H distance in these geometries is 1.5 \AA. The fitting curves (orange and green dashed lines) 
    are line functions, $y=kx+c$, where $k$ and $c$ are fitting coefficients. (c) The $N_{\rm Unitaries}/N_{\rm Paulis}$ ratios 
    for the Hamiltonians of a wider class of quantum molecular systems encoded using up to 40 qubits. 
    Ratios for a series of 20 qubit systems studied are shown in histograms in the inset. 
    The fitting curves take the expression, $\log y = a \log x + b$, where $a$ and $b$ are fitting coefficients.}
    \label{fig:unitary_search}
\end{figure*}

%Typically, an arbitrary non-unitary operator (mainly the Fermionic operator in the present discussion) 
%can be transformed into a linear combination of Pauli terms. Since each Pauli term is a unitary operation, 
%this is the basic version of a linear combination of unitaries. If considering a fault-tolerant implementation 
%with favorable resource scaling, a general non-unitary operator can be encoded through a linear combination of 
%unitaries technique,\cite{Childs2012} amplitude amplification approach,\cite{brassard2002quantum}  
%Hamiltonian simulation,\cite{Berry2015Simulating} qubitization,\cite{low2017optimal} or the direct 
%block-encoding\cite{gilyen2019quantum} methods at the cost of introducing deeper circuit and implementing 
%controlled-$\hat{\mathcal{U}}$ operations, which nevertheless come with a probability of failure and 
%require advanced circuit and error mitigation that might go beyond the capability of the current NISQ devices. 
%Alternatively, Izmaylov et al. proposed a unitary partitioning approach \cite{Izmaylov2019} for 
%encoding a linear combination of Pauli terms on a quantum circuit. Here we  briefly review this approach 
%and then propose a more efficient unitary partitioning approach guided by the SR used in the simulation. 

The original version of the unitary partitioning approach proposed in Ref.  \citenum{Izmaylov2019} 
aims to search a limited number of unitaries, $\{ U_1, U_2, \cdots, U_L \}$, 
whose linear combination is equivalent to the linear combination of original Pauli terms, $\{ P_1, P_2, \cdots, P_S \}$, i.e.,
\begin{align}
   \sum_{l=1}^L u_l U_l = \sum_{s=1}^S p_s P_s, ~~ L < S. \label{findU}
\end{align}
To see how it can be done, we use a simple example
to combine two Pauli terms $p_1 P_1 + p_2 P_2$ ($p_1$ and $p_2$ are scalar coefficients) to form a unitary $U_l$ multiplied by a scalar coefficient $u_l$. The unitarity requires
\begin{align}
    |u_l|^2 I &= (p_1^\ast P_1 + p_2^\ast P_2) (p_1 P_1 + p_2 P_2) \notag \\
     &= (|p_1|^2 + |p_2|^2) I + (p_1^\ast p_2 P_1 P_2 + p_2^\ast p_1 P_2 P_1) \label{unitary_cond}
\end{align}
For arbitrary $P_1$ and $P_2$, one sufficient condition to hold Eq. (\ref{unitary_cond}) is an anticommutation-type relation between $P_1$ and $P_2$,
\begin{align}
    p_1^\ast p_2 P_1 P_2 + p_2^\ast p_1 P_2 P_1 = 0. \label{U_cond1}
\end{align}
For the unitarization of a linear combination of more than two Pauli terms, Eq. (\ref{U_cond1}) can be generalized as
\begin{align}
    \sum_{r,s=1,r\neq s}^S p_s^\ast p_r P_s P_r = 0, ~~ S > 2,
\end{align}
which can be satisfied if any two Pauli terms from the string list satisfy Eq. (\ref{U_cond1}). 
These sufficient conditions then translate searching $\{U_l\}$ satisfying Eq. (\ref{findU}) 
to a graph problem, where all the weighted Pauli terms are taken as the nodes of a graph $G$, 
and an edge can be drawn between two nodes if they satisfy Eq. (\ref{U_cond1}). 
Thus, a unitary can be found for a number of Pauli terms if their corresponding nodes form a clique. 
A search of a minimal number of $U_l$'s for a set of Pauli terms $\{P_s\}$ weighted by 
the coefficients $p_s$'s is then translated to a graph ``minimal clique cover" problem, 
which is NP-hard in general, but can be approximately solved by polynomial approaches 
such as the graph coloring algorithms for the complement graph of $G$. A schematic representation of the unitary partitioning approach is presented in Fig. \ref{fig:unitary_search}a. The robustness of unitary partitioning for some molecular Hamiltonians have already been demonstrated in Ref.  \citenum{Izmaylov2019}, where the number of unitary terms is roughly one order of magnitude smaller than the number of Pauli terms generated from either Jordan-Wigner\cite{JW1928} or Bravyi-Kitaev\cite{bravyi2002Fermionic} transformations. 
In Fig. \ref{fig:unitary_search}b and c, 
we verified the performance of this approach using a wider class of molecular 
Hamiltonians associated with $\sim$40 molecules with varying basis sets. 
These Hamiltonians are encoded using 4$\sim$40 qubits. As can be seen in Fig. \ref{fig:unitary_search}b, 
for the Hamiltonians of H$_n$ ($n=2,4,6,8,10$) 1D/2D/3D systems, the number of the generated unitaries, 
$N_{\rm Unitaries}$, scales as $\mathcal{O}(N_{\rm Qubits}^{2.59})$ (with $N_{\rm Qubits}$ the number of qubits), 
and the order is about $\sim$1.22 smaller than the scaling order of the number of Pauli terms, 
$N_{\rm Paulis}$. In Fig. \ref{fig:unitary_search}c, the ``compression" ratios, defined by 
$N_{\rm Unitaries}/N_{\rm Paulis}$, drops significantly, as implied by the fitting curve, 
as an inverse function of $N_{\rm Qubits}$ with the order roughly $\sim$-1.27. 
Also, from the inset of Fig. \ref{fig:unitary_search}c, even though the 20-qubit category 
includes a wider range of molecular Hamiltonians, the ``compression" ratios for all the 
20-qubit systems are quite consistent at $\sim$0.1. It is worth noting that similar 
algorithms with different definitions of the ``edge" in the graph have routinely been employed 
for partitioning the $\mathcal{O}(N^4)$ terms in the Hamiltonian into the qubit-wise commutative groups for simultaneous measurements.\cite{Kandala2017hardware,mcclean2016theory,Izmaylov2020} 
\textcolor{black}{When approaching the local basis limit, since the electron-electron term goes 
asymptotically to $\mathcal{O}(N^2)$, the Hamiltonian will be approaching to a diagonal form in the limit, 
and a normalized diagonal matrix can be formally represented as the sum of two unitaries. 
For conventional cases, applications of other grouping techniques \cite{gokhale2020,Izmaylov2020,Izmaylov2019,Huggins_2021} 
for evaluating $\langle H \rangle$ have shown that, at a cost of introducing 
additional one-/multi-qubit unitary transformations before the measurement, 
the total number of terms can be significantly reduced from $\mathcal{O}(N^4)$ to $\mathcal{O}(N^{2\sim3})$.}

Despite the success in partitioning Hamiltonians represented in Pauli terms, the straightforward application of this unitary-partitioning approach for representing other non-unitary operators, including the CC exponential operator for ket and bra states, needs to be scrutinized. 
Another class of challenges is associated with approximating $|\Omega\rangle$ (as well as in  $|\Gamma\rangle$) vector and the form of the $e^{T^{(A)}}|\Phi\rangle$ expansion.
As discussed in Ref. \citenum{jankowski1991method}, the length of this finite expansion depends 
on the total number of correlated particles. 
Though, as a consequence of nilpotent character of the operator algebra used to define the excitations in cluster operator,  
it is possible to build exact expansion for $e^{T^{(A)}}|\Phi\rangle$; in this paper, 
we consider simple approximations of its form that can reproduce leading contributions to the corresponding moments. 
Take the CCSD wave functions as examples. 
\textcolor{black}{Employing the bi-variational CC formulation with singles and doubles, ($\mathcal{L}$-)CCSD, where $T=T_1+T_2$ and the linear expansion of $\mathcal{L}$, Eq. (\ref{L_T}), is truncated at the second order (i.e., $\mathcal{L}=1+\Lambda_1+\Lambda_2$), }
%
%If we only focus on reproducing the CCSD energy, 
the CCSD bra and ket states can then be approximately expanded as
\begin{align}
    \langle \Psi_{\rm CCSD}| 
    &= \langle \Phi | \mathcal{L} e^{-T_1-T_2} \notag \\
    &\approx \langle \Phi | (1 + \Lambda_1 + \Lambda_2 )( 1 - T_1 - T_2 + \frac{1}{2}T_1^2), \label{bra_expansion} \\ 
    |\Psi_{\rm CCSD}\rangle 
    &= e^{T_1+T_2} |\Phi \rangle \notag \\
    &\approx (1 + T_1 + T_2 + \frac{1}{2}T_1^2 + T_1T_2 + \frac{1}{6}T_1^3 + \frac{1}{2}T_2^2 \notag \\
    &~~~~ + \frac{1}{24}T_1^4) | \Phi \rangle . \label{ket_expansion}
\end{align}
from which the CCSD energy can be computed through
\begin{align}
    E_{\rm{CCSD}} &= \langle \Phi | H e^{T_1+T_2} | \Phi \rangle \notag \\
    &= \langle \Phi | \mathcal{L} e^{-T_1-T_2} H e^{T_1+T_2} | \Phi \rangle \label{e_ccsd}
\end{align}
The unitary partitioning can directly be applied to the Pauli terms generated for the 
right-hand side of Eqs. (\ref{bra_expansion}) and (\ref{ket_expansion}) 
using typical Jordan-Wigner,\cite{JW1928} Bravyi-Kitaev,\cite{bravyi2002Fermionic} 
or parity transformation.\cite{tranter2015Bravyi} 
However, from Tab. \ref{tab:comparison}, the number of Pauli terms generated from 
these typical transformations for either CCSD bra or ket states is approximately 
represented by Eqs. (\ref{bra_expansion}) and (\ref{ket_expansion}) and seems to grow exponentially, 
e.g., from the H4 1D/2D/3D systems encoded using 8 qubits to the H6 1D/2D/3D systems encoded using 12 qubits. 
The numbers of the associated Pauli terms are increased by roughly two orders of magnitude. 
The unitary partitioning approach can only moderately ``compress" the numbers of Pauli terms by $\frac{1}{6}\sim\frac{1}{2}$, which are still large numbers for viable quantum implementation. 

To tackle this hindrance, we propose a modified unitary partitioning approach for 
formally and efficiently encoding the non-unitary ansatz, such as the conventional CC ansatz 
on quantum devices. Our proposed approach can be referred to as a SR-guided unitary approach, 
which offers a more compact unitary basis that has equivalent effect as the non-unitary operators 
when acting on a chosen reference state $|\Phi\rangle$, and is detailed below.

For the convenience of discussion, throughout this paper we label the occupied 
spin-orbitals by the indices $i,j,k,\dots$, and the virtual spin-orbitals by the indices $a,b,c,\dots$. We start by noticing that for a single-determinant reference state  $|\Phi\rangle = |\cdots j \cdots i\rangle$ since 
\begin{align}
a_i^\dagger |\Phi\rangle = 0, ~~~~
a_a |\Phi\rangle = 0, \label{cond_single_det}
\end{align}
the following equivalencies hold
\begin{align}
    a_i |\Phi\rangle = \left\{ \begin{array}{l}
    (a_i + a_i^\dagger) |\Phi\rangle \\
    (a_i - a_i^\dagger) |\Phi\rangle 
    \end{array} \right., ~
    a_a^\dagger |\Phi\rangle = \left\{ \begin{array}{l}
    (a_a^\dagger + a_a) |\Phi\rangle \\
    (a_a^\dagger - a_a) |\Phi\rangle 
    \end{array} \right. \label{op}.
\end{align}
If we map the indexed fermionic operators to the corresponding qubit operators (via Jordan-Wigner transformation) as follows
\begin{align}
\left\{ \begin{array}{rl}
    a_p^\dagger & = \frac{1}{2} Z_0 \cdots Z_{p-1} (X_p - iY_q) \\
    a_p         & = \frac{1}{2} Z_0 \cdots Z_{p-1} (X_p + iY_p)
\end{array}\right. , \label{JW_trans}
\end{align}
Eqs. (\ref{op}) can then be transformed as
\begin{align}
    &a_i |\Phi\rangle = \left\{ \begin{array}{r}
    ~~~Z_0\cdots Z_{i-1} X_i |\Phi\rangle \\
    ~~~i Z_0\cdots Z_{i-1} Y_i |\Phi\rangle
    \end{array} \right. , \label{anni_q} \\
    &a_a^\dagger |\Phi\rangle = \left\{ \begin{array}{r}
    Z_0\cdots Z_{a-1} X_a |\Phi\rangle \\
    -i Z_0\cdots Z_{a-1} Y_a |\Phi\rangle 
    \end{array} \right. \label{crea_q}.
\end{align}
\textcolor{black}{Note that in comparison to Eq. (\ref{JW_trans}), which transforms one creation/annihilation 
operation to two Pauli strings, Eqs. (\ref{anni_q}) and (\ref{crea_q}) transform one creation/annihilation 
operation only to one Pauli string under the condition of Eq. (\ref{cond_single_det}) that $|\Phi\rangle$ 
needs to be a single determinant.}
Now we can utilize Eqs. (\ref{anni_q}) and (\ref{crea_q}) to map an arbitrary 
Fermonic operator acting on the reference to a unitary operator acting on the same reference. 
Take the conventional CC single and double wave functions 
\begin{align}
    e^T |\Phi\rangle &\approx e^{T_1+T_2} |\Phi\rangle \notag \\
    &= \left( 1 + \sum_{i,a}t_i^a a_a^\dagger a_i + \sum_{\substack{i<j\\a<b}} \tilde{t}_{ij}^{ab} a_a^\dagger a_b^\dagger a_j a_i + \cdots \right) |\Phi\rangle 
    \label{ccwfn}
\end{align} 
(with $\tilde{t}_{ij}^{ab} = t_{ij}^{ab} + t_i^a t_j^b - t_i^b t_j^a$) as an example, we can do the following transformation for the single and double terms
\begin{align}
t_i^a a_a^\dagger a_i |\Phi\rangle 
&= t_i^a \left( \begin{array}{lll}
    \cdots \otimes X_a &\cdots \otimes  Z_i &\cdots  \\
    \cdots \otimes I &\cdots \otimes  X_i  &\cdots 
    \end{array} \right) \Bigg| \Phi \Bigg\rangle \notag \\
&= i t_i^a X_a \cdots Z_m \cdots Y_i |\Phi\rangle, ~~~~  i < m < a, \label{single}
\end{align}
\begin{align}
    &\tilde{t}_{ij}^{ab} a_a^\dagger a_b^\dagger a_j a_i |\Phi\rangle \notag \\
    &~= \tilde{t}_{ij}^{ab} \left(
    \begin{array}{lllll}
    \cdots \otimes I &\cdots \otimes X_a &\cdots \otimes Z_j &\cdots \otimes Z_i &\cdots  \\
    \cdots \otimes X_b &\cdots \otimes Z_a &\cdots \otimes Z_j &\cdots \otimes Z_i &\cdots  \\
    \cdots \otimes I &\cdots \otimes I &\cdots \otimes  X_j  &\cdots \otimes Z_i &\cdots  \\
    \cdots \otimes I &\cdots \otimes I   &\cdots \otimes  I  &\cdots \otimes X_i &\cdots 
    \end{array}   
   \right) \Bigg| \Phi \Bigg\rangle \notag \\
   &~= \tilde{t}_{ij}^{ab} X_b \cdots Z_c \cdots Y_a X_j \cdots Z_m \cdots Y_i |\Phi\rangle, ~~~~ 
   \begin{array}{c}
    i < m < j    \\  a < c < b
   \end{array},
   \label{double}
\end{align}
thus Eq. (\ref{ccwfn}) can be represented by a linear combination of Pauli terms 
%without explicitly performing standard encodings such as Jordan-Wigner, Bravyi-Kitaev, or parity transformations, 
\textcolor{black}{by directly applying Eqs. (\ref{anni_q}) and (\ref{crea_q}),} 
and the number of Pauli terms to represent the single and double terms in the CCSD wave function 
scales as $\mathcal{O}(N_oN_v)$ and $\mathcal{O}(N_o^2N_v^2)$, respectively, with $N_o$ the number 
of occupied spin-orbitals and $N_v$ the number of virtual spin-orbitals. We then apply the unitary 
partitioning approach on these new Pauli terms aiming for a more compact unitary representation for 
the CCSD bra and ket states. Some preliminary results are shown in Tab. \ref{tab:comparison} (more 
extensive scaling analysis is given in the following numerical section). As can be seen, for the 
8-qubit H4 1D/2D/3D systems, the number of unitaries generated from this SR-guided 
unitary partitioning is about 5$\sim$12, in comparison to 200$\sim$400 using the unitary partitioning 
on the original Pauli terms. For slightly larger 12-qubit H6 1D/2D/3D systems, the reduction is 
even more significant, i.e., $10\sim 30$ versus $1\times 10^4\sim 5 \times 10^4$.

\begin{table*}[!htbp]
\resizebox{\linewidth}{!}{%
\begin{tabular}{llccccccccccccc}
\hline
                & \multirow{2}{*}{Method}   & \multicolumn{6}{c}{$\langle \Psi_{\rm CCSD}|$} & &\multicolumn{6}{c}{$| \Psi_{\rm CCSD} \rangle$}\\  \cline{3-8} \cline{10-15}
                &   & H4 (1D)   & H4 (2D)   & H4 (3D)   & H6 (1D)   & H6 (2D)   & H6 (3D)   && H4 (1D)    & H4 (2D)    & H4 (3D)   & H6 (1D)  & H6 (2D)   & H6 (3D)\\ \hline
$N_{\rm Paulis}$ & JW/BK/Parity   & 881       & 553       & 1397      & 27041     & 46585 & 12857 && 689        & 457        & 1057      & 19025     & 28441   & 9897 \\
$N_{\rm Unitaries}$ & UP & 315       & 233       & 217       & 7661      & 3476  & 1148  && 268        & 181        & 393       & 5771      & 8229    & 2858 \\
$N_{\rm Paulis}$ & S-JW (this work)        & 15        & 7         & 27        & 60        & 98    & 60    && 20         & 10         & 36        & 191       & 345     & 191 \\
$N_{\rm Unitaries}$ & UP (this work)    & 11        & 5         & 5         & 22        & 18    & 11    && 12         & 6          & 6         & 33        & 33      & 20\\ \hline
\end{tabular}}
\caption{The number of Pauli terms, $N_{\rm Paulis}$, and the number of the generated unitaries, $N_{\rm Unitaries}$, 
representing the CCSD bra and ket states for the H4 (1D/2D/3D) and H6 (1D/2D/3D) systems. 
The geometries of the H4 and H6 systems are given in the supplementary materials. 
Original Pauli terms are generated using conventional Jordan-Wigner (JW),\cite{JW1928} Bravyi-Kitaev (BK),
\cite{bravyi2002Fermionic} or parity transformation,\cite{tranter2015Bravyi} 
and we found these transformations generated the same number $N_{\rm Paulis}$ for the studied H4 and H6 systems. 
We also exhibit the number of Pauli terms generated using the SR-guided scheme combining with the conventional 
Jordan-Wigner transformation, denoted as ``S-JW", described in Section \ref{SRGUP}. 
The unitaries are then generated using the unitary partitioning (UP) approach on the original Pauli terms and 
the SR-guided Pauli terms, respectively. 
\textcolor{black}{The $T_1$ and $T_2$ ($\Lambda_1$ and $\Lambda_2$) amplitudes are computed 
from solving the conventional CCSD single and double residual equations employing a classical computer, 
where the STO-3G basis set\cite{Stewart1970Small} is employed for all the calculations. 
The converged $T_1$ and $T_2$  ($\Lambda_1$ and $\Lambda_2$) amplitudes are then plugged into 
Eqs. (\ref{bra_expansion}) and (\ref{ket_expansion}) to form the approximate Fermionic forms of 
the CCSD left and right wave functions, i.e., $\langle \Psi_{\rm CCSD}|$ and $| \Psi_{\rm CCSD} \rangle$.}
%The STO-3G basis set\cite{Stewart1970Small} is employed for all the classical CCSD calculations to get the converged CCSD $T_1/T_2$ amplitudes. $\langle \Psi_{\rm CCSD}|$ and $| \Psi_{\rm CCSD} \rangle$ are approximately expanded according to Eqs. (\ref{bra_expansion},\ref{ket_expansion}).
}
\label{tab:comparison}
\end{table*}

\section{Numerical examples}
\label{section5}

In this section, we provide examples illustrating the performance of our proposed unitary partitioning techniques 
and their applications in quantum computing, and the basic variants of MMCC corrections, 
including $\Lambda$-based MMCC corrections (see Eq. (\ref{eq8})) and those used in the early stages of 
validating the renormalized approaches on classical computers. 

\subsection{Performance of SR-guided unitary partitioning over a wide class of molecular systems}

%The core of our proposed quantum algorithm is the search of a compact unitary basis through graph analysis of Pauli terms and their anti-commuting relationship. 
To give a performance analysis of our proposed SR-guided unitary paritioning approach, 
we chose a wide class of molecular systems (geometries are given in the supplementary materials), including 
\textcolor{black}{13 hydrogenic molecules consisting of one H$_2$ molecule and 12 H$n$ ($n=4,6,8,10$) hydrogenic molecules in 1D/2D/3D, as well as 23} 
simple molecules composed of the first- and second-row elements from the periodic table, and using 4$\sim$40 qubits to encode their %corresponding electronic configurations, 
\textcolor{black}{initial Hartree-Fock states, }
to test its robustness for representing the corresponding CCSD bra and 
ket states (Eqs. (\ref{bra_expansion}) and (\ref{ket_expansion})). %reproduce the CCSD energies,
\textcolor{black}{The stability of our approach is tested by employing our quantum approach 
to reproduce the CCSD energies of the 13 hydrogenic molecules as well as LiH, HF, BeH$_2$, and H$_2$O molecules. 
The computed energies from a classical CCSD approach and our quantum approach employing 
quantum simulators are shown in Appendix \ref{app_table}, Tab. \ref{tab:comparison1}. }

\begin{figure*}
    \centering
    \includegraphics[width=\linewidth]{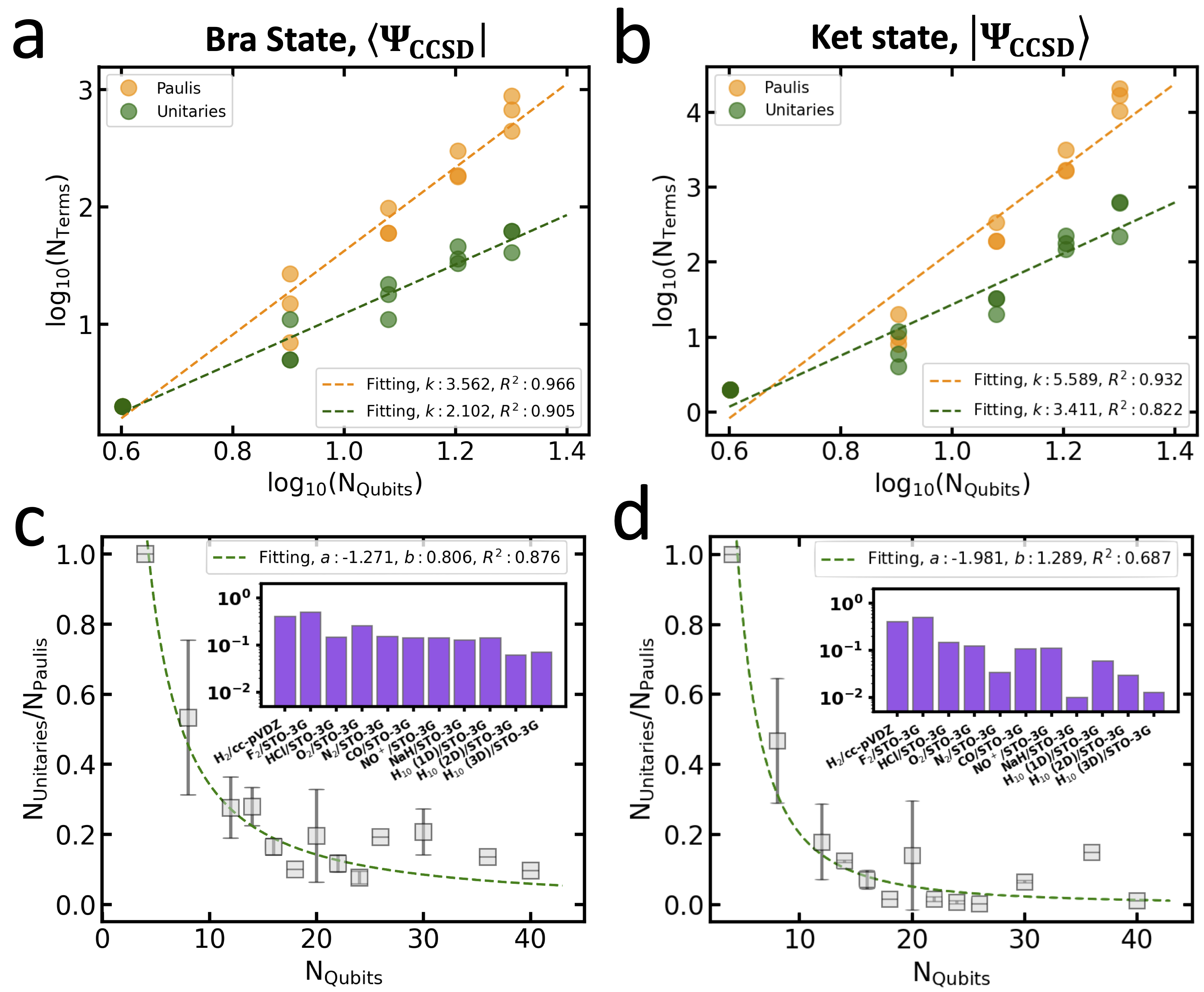}
    \caption{(a,b) Numbers of Pauli terms ($N_{\rm Paulis}$) and corresponding numbers of unitary groups ($N_{\rm Unitaries}$) determined from the SR-guided unitary partitioning approach for the approximate expansion of $\langle \Psi_{\rm CCSD}|$ and $|\Psi_{\rm CCSD}\rangle$ of 1D, 2D, and 3D hydrogen systems encoded using up to 20 qubits. The expansions of the left and right CCSD wave functions are shown in Eqs. (\ref{bra_expansion}) and (\ref{ket_expansion}). In these hydrogen systems, the nearest H$-$H distance is 1.5 \AA. The fitting curves (orange and green dashed lines) are line functions, $y=kx+c$, where $k$ and $c$ are fitting coefficients. (c,d) The $N_{\rm Unitaries}/N_{\rm Paulis}$ ratios for $\langle \Psi_{\rm CCSD}|$ and $|\Psi_{\rm CCSD}\rangle$ of a wider class of quantum molecular systems encoded using up to 40 qubits. Ratios for a series of 20 qubit systems studied are shown in histograms in the insets. The fitting curves take the expression, $\log y = a \log x + b$, where $a$ and $b$ are fitting coefficients.}
    \label{fig:ref_unitary_search}
\end{figure*}

The results are shown in Fig. \ref{fig:ref_unitary_search}. We first tested a series of 1D, 2D, and 3D hydrogenic molecules
%(totally 13 Hydrogenic molecules) consisting of up to 10 Hydrogen atoms 
described by the STO-3G basis set. The scaling of the number of generated unitaries, 
in comparison with the scaling of the number of the Pauli terms 
(generated by the SR-guided Jordan-Wigner transformation, or S-JW), 
for representing the CCSD bra and ket states are exhibited in Fig. \ref{fig:ref_unitary_search}a and b. 
As can be seen for these hydrogen systems, the scalings of the number of generated unitaries 
as functions of the number of qubits are about 1.46$\sim$2.17 order of magnitude lower than those of the Pauli terms. 
The extended analysis of the $N_{\rm Unitaries}/N_{\rm Paulis}$ ``compression" ratios as 
functions of the number of qubits on a wider class of 41 simple molecular systems in 
Fig. \ref{fig:ref_unitary_search}c and d have shown that the ratio drops significantly 
when approaching the larger number of qubits, and the fitting curves imply the ratio drops 
as an inverse function of the number of qubits with the inverse power roughly -1.27 and -1.98 for 
the CCSD bra and ket non-unitary operator, respectively. Also, despite the ``compression" ratios,  
all 20 qubit systems (insets of Fig. \ref{fig:ref_unitary_search}c and d) exhibit a 
slightly wider variance, and the ratio is well below 40\% even in the worst cases (e.g., F$_2$/STO-3G and H$_2$/cc-pVDZ). 
The CCSD energies computed through measuring the unitary products have been examined to 
fully reproduce classical CCSD energies for all the hydrogen systems with STO-3G basis, 
together with some other small molecules (see tabulated data in supplementary materials). 

\textcolor{black}{The CCSD energies computed from both classical and our proposed quantum approaches, 
as shown in Appendix \ref{app_table}, Tab. \ref{tab:comparison1}, indicate that Eqs. (\ref{bra_expansion}) and (\ref{ket_expansion}) 
are sufficient for reproducing the CCSD energy with less than micro-Hartree energy difference for the studied test set. 
One can definitely use larger expansion for higher order CC approaches (see the following sections). 
Although the number of Pauli terms might increase substantially in high-order CC approaches, 
$N_{\rm Unitaries}$ would be reaching one when the CC wave functions approach the exact wave function limit. 
This is because the limit could also be achieved by the unitary CC form.}
It is worth noting that our proposed unitary partitioning scheme also works for the product of non-unitary operators. 
Take the $He^T|\Phi\rangle$ for an example, one can either do the 
unitary partitioning for $H$ and $e^T|\Phi\rangle$ separately, 
then combine the unitaries to get the compact unitary representation of $He^T|\Phi\rangle$, 
or if some extra classical computation becomes affordable for obtaining the Pauli terms for 
the combined $He^T|\Phi\rangle$, perform a one-shot unitary partitioning directly on the SR-guided 
Pauli terms of $He^T|\Phi\rangle$. Usually, the second route would generate a more compact unitary representation 
than the first. For example, for H6 1D/2D/3D systems with STO-3G basis, switching from the first route to 
the second can significantly reduce the number of unitaries that needs to be prepared 
from $174\times33$, $155\times33$, and $93\times20$ (i.e. $N_{\rm Unitaries}^{H}\times N_{\rm Unitaries}^{|\Psi_{\rm CCSD}\rangle}$) to 1010, 847, and 290 (i.e. $N_{\rm Unitaries}^{H|\Psi_{\rm CCSD}\rangle}$) with the ``compression" ratios roughly $\frac{1}{6}$.

\subsection{MMCC quantum simulation of single-impurity Anderson model}

With our proposed SR-guided unitary partitioning approach, we are now able to target non-unitary calculations 
through quantum simulations. Our first demonstration is to prepare a conventional CC ansatz for a half-filled, four-site, 
single impurity Anderson model (SIAM) system. The general SIAM Hamiltonian reads
\begin{align}
H_{\rm SIAM} = H_{\rm imp.} + H_{\rm bath} + H_{hyb.}, \label{Hsiam}
\end{align}
where 
\begin{align}
H_{\rm imp.} = \sum_{\sigma} \epsilon_c c^\dagger_{\sigma} c_{\sigma} + U c^\dagger_{\uparrow}c_{\uparrow}c^\dagger_{\downarrow}c_{\downarrow} \label{Himp}
\end{align}
describes the impurity-site energy $\epsilon_c$ and the Coulomb interaction $U$ between the electrons with opposite spins ($\sigma = \uparrow$ or $\downarrow$) at the impurity site,
\begin{align}
H_{\rm bath} = \sum_{i=1,\sigma}^{N_{\rm b}} \epsilon_{d,i} d^\dagger_{i,\sigma} d_{i,\sigma} \label{Hbath}
\end{align}
describes the non-interacting bath site with $\epsilon_{d}$ being the bath-site energies, and
\begin{align}
H_{\rm hyb.} = \sum_{i=1,\sigma}^{N_{\rm b}} V_i \big( c^\dagger_{\sigma} d_{i,\sigma} + d^\dagger_{i,\sigma} c_{\sigma} \big) \label{Hhyb}
\end{align}
describes the coupling between the impurity site and the bath levels due to hybridization. 
Here, we only consider symmetric four-site SIAM with $\epsilon_c = -\frac{U}{2}$. 
For the four-site SIAM, there are a total of eight spin-orbitals, 
which are then mapped to eight qubits for quantum simulation. 
The initial electronic configuration is $|10101010\rangle$. In the present demonstration, we set the bath orbital energies to be evenly distributed between $\left[-1,1\right]$, and assume homogeneous coupling with $V = 1$. In the meanwhile, we select six $U$ values such that the corresponding $\log_{10}(U/V)$ values are evenly distributed within $\left[0.0,1.5\right]$. The quantum simulations for evaluating the MMCC energies of the system are focused on measuring the expectation values of the combined unitaries generated from our proposed unitary partitioning scheme on SIAM Hamiltonian and its corresponding CCSD bra and ket states with respect to the reference state.

Given the size of the select SIAM, following MMCC methodology, 
we only consider including triples in the left and right CC entanglers to constitute 
higher order correction to the CCSD energy, e.g., CCSDT energy. Between CCSD and CCSDT, 
we also consider one other energy functional in the MMCC framework, the CCSD-$\Lambda_3$ functional, 
generated by including the converged $\Lambda_3$ from the $\Lambda$-CCSDT calculations into $\mathcal{L}$, i.e.,
\begin{align}
    \mathcal{L} &\approx 1 + \Lambda_1 + \Lambda_2 + \Lambda_3,
\end{align}
and keeping the expansions of $e^{\pm T}$ unchanged.

\begin{figure*}
    \centering
    \includegraphics[width=\linewidth]{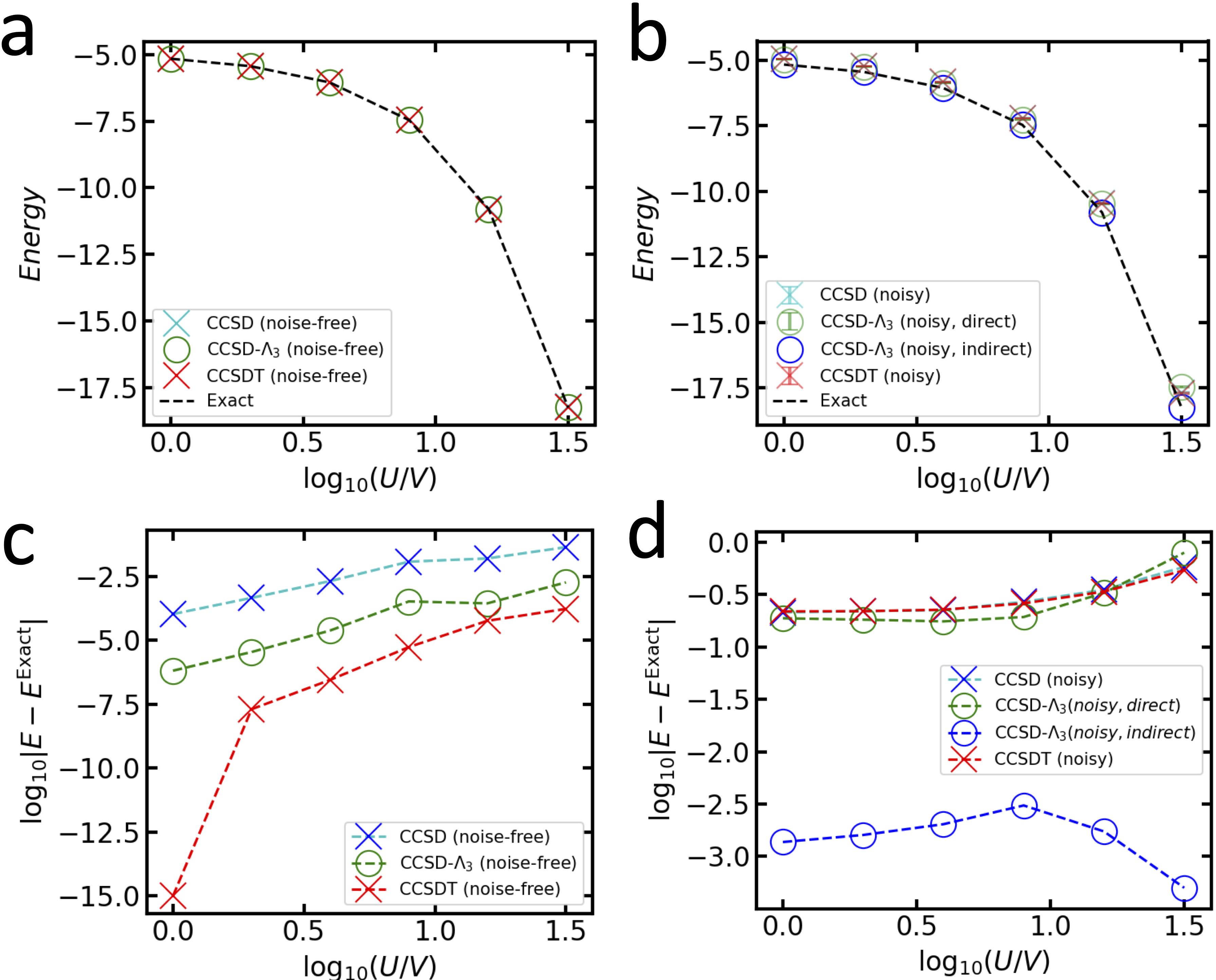}
    \caption{MMCC energy estimates from noise-free (a,c) and noisy (b,d) quantum simulations of the ground-state energy of 
    the half-filling, four-site SIAM at different $U/V$ ratios \textcolor{black}{(here we set $V=1.0$ 
    and vary $U$ values to get different $U/V$ ratios)}. The noisy quantum simulations using direct 
    and indirect measurements are performed with IBM Qiskit employing the \textit{Fake\_Toronto} 
    quantum simulator that mimics the  simulation on the real \textit{ibmq\_toronto} quantum device. 
    Each noisy data point and the associated error bar in (c) represents an average measurement outcome 
    and the associated standard deviation from 10 experiments with 10,000 shots for each experiment. 
    Tabulated data are provided in the supplementary materials.}
    \label{fig:siam_results}
\end{figure*}

Fig. \ref{fig:siam_results} shows the computed CC energies of the four-site SIAM at different $U/V$ values. 
The employed CC wave function methods exhibit improved energy estimates from CCSD to CCSD-$\Lambda_3$ to CCSDT 
over the entire range of the studied $U/V$ ratios, with more significant improvement observed at smaller $U/V$ 
ratios associated with weaker correlation. We also perform MMCC quantum simulations using the noisy simulator. 
As shown in Fig. \ref{fig:siam_results}b and d, in the noisy simulation, the CCSD energies slightly but visibly deviate 
and a slightly larger deviation can be observed for larger $U/V$ ratios. 
It is worth mentioning that the CCSD-$\Lambda_3$ energies can also be 
computed through indirect measurement (blue circles in Fig. \ref{fig:siam_results}b and d), 
where we only measure the unitary products that contribute to the expectations 
$\langle \Phi |\Lambda_3 e^{-T}|H|e^T|\Phi\rangle$ and $\langle \Phi |\Lambda_3 e^{-T}|\cdot|e^{T}|\Phi\rangle$, 
assume an accurate $E_{\rm CCSD}$, and resemble $E_{\rm CCSD-\Lambda_3}$ through 
\begin{align}
E_{\rm CCSD-\Lambda_3} &= \frac{\langle \Psi_T|H|e^{T}|\Phi\rangle}{\langle\Psi_T|e^{T}|\Phi\rangle} \notag \\
&= \frac{E_{\rm CCSD}+\langle \Phi |\Lambda_3 e^{-T}|H|e^T|\Phi\rangle}{1 + \langle \Phi |\Lambda_3 e^{-T}|\cdot|e^{T}|\Phi\rangle}.
\end{align}
Since the indirect measurement mitigates the error from $E_{\rm CCSD}$, 
the entire error is thus removed to a large extent, 
and the simulation only focuses on a relatively small energy difference, 
which explains why the deviation with reference to the exact results are reduced by two to three orders of magnitude, 
in particular for larger $U/V$ ratios. 
\textcolor{black}{It is worth mentioning that the relatively big deviation in the strongly correlated regime 
(i.e., larger U/V ratios) shown in Fig. \ref{fig:siam_results} is due to the CC method employed, 
which is independent of the proposed unitary partitioning approach, 
a generic approach that can be used to reproduce any non-unitary results with the quality 
of the results solely dependent on non-unitary method itself. This has been well demonstrated 
in Fig. \ref{fig:siam_results}c, where in a larger U/V regime, with or without unitary partitioning, 
the CCSDT results outperform the CCSD methods (for large $U/V$ ratios, 
the absolute deviation of the CCSDT results with respect to the exact results is $<10^{-3}$), 
and our unitary representation is able to reproduce both CCSD and CCSDT results.}

\begin{figure*}
    \centering
    \includegraphics[width=\linewidth]{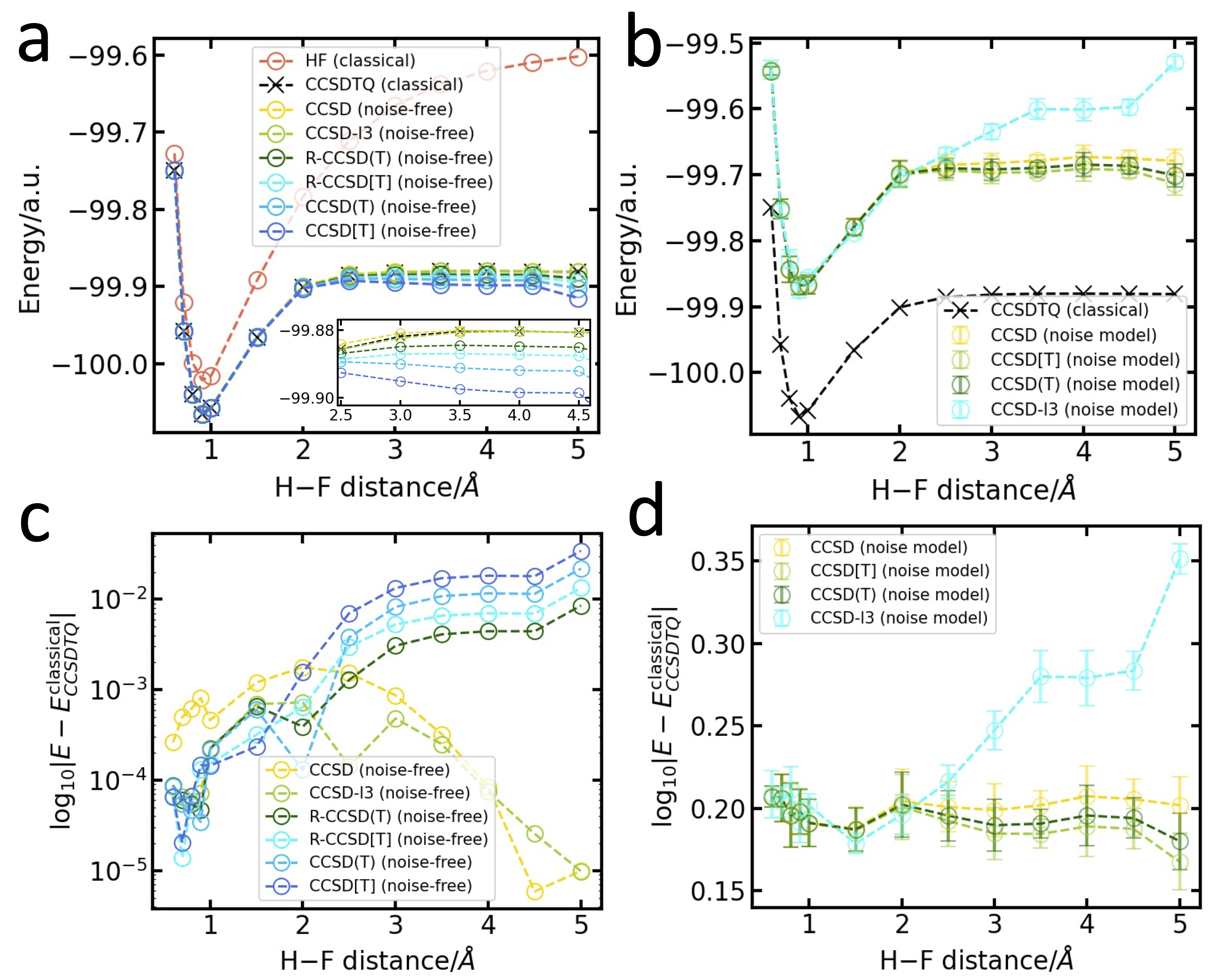}
    \caption{The PES of the HF molecule computed from noise-free simulation (a,c) and noisy simulation (b,d) 
    employing different CC approaches. The double-zeta basis is employed for all  quantum simulations 
    in which the active space $(6o,6e)$ is chosen to map to 12 qubits. 
    \textcolor{black}{For the HF molecule with the (6o,6e) active space, 
    the CCSDTQ results are very close to the exact full configuration interaction solution.} 
    The noisy quantum simulations are performed using IBM Qiskit employing the \textit{Fake\_Toronto} 
    quantum simulator that mimics the simulation on the real \textit{ibmq\_toronto} quantum device. 
    Each CC data point and the associated error bar in (b,d) represents an average measurement outcome 
    and the associated standard deviation from 10 experiments with 1,024 shots for each experiment. 
    Tabulated data are provided in the supplementary materials.}
    \label{fig:HF_results}
\end{figure*}

\subsection{MMCC quantum simulation of hydrogen fluoride (HF) molecule}

With the proposed unitary partitioning approach, we are also able to quantum simulate 
the PES of molecular systems employing the MMCC approaches. To demonstrate the capability, 
we choose the HF molecule described by the double-zeta basis set. The HF molecular systems 
have been extensively studied using the MMCC approaches in  classical computing 
(for example, see Refs. \citenum{mmcc1} and \citenum{mmcc2}), where it has been found that 
the effect of the connected tetraexcited clusters to the energy correction is rather small, 
thus $T_3$ clusters play an important role in energy correction. Our quantum simulations 
correspond to six classical CC energy expressions, namely CCSD, CCSD-l3, 
(R-)CCSD[T],\cite{mmcc1} and (R-)CCSD(T) \cite{mmcc1} (`R' is short for renormalized), 
whose results are then compared with the Hartree-Fock and classical CCSDTQ results 
(the latter is very close to the exact solution in the full molecular orbital space, 
see Refs. \citenum{mmcc1} and \citenum{mmcc2}). Among these MMCC energy expressions, 
the CCSD-l3 expression is the same as that of CCSD-$\Lambda_3$ except that a perturbative approximation 
is introduced for $\Lambda_3$ (\textcolor{black}{see Ref. \citenum{mmcc1}}), i.e.,
\begin{align}
    \Lambda_3 \approx (R_0^{(3)}(V_NT_2)_C)^\dagger
\end{align}
where $T_2$ is obtained in the CCSD calculations, $V_N$ is the two-body part of the Hamiltonian 
in the normal product form ($H_N$), and $R_0^{(3)}$.\textcolor{black}{designates three-body zeroth-order resolvent, 
i.e., the Hamiltonian in the standard resolvent operator is approximated by the Fock operator.}
%sum of differences of orbital energies,
%$\epsilon_i+\epsilon_j+\epsilon_k-\epsilon_a-\epsilon_b-\epsilon_c$
%designates the three-body part of the reduced resolvent from many-body perturbation theory (MBPT). 
The CCSD[T] and CCSD(T) energy corrections are computed through
\begin{align}
    \Delta E_{\rm CCSD[T]/CCSD(T)} &= \langle \Psi_{\rm CCSD[T]/CCSD(T)} | Q_3(V_NT_2)_C|\Phi\rangle
\end{align}
where $Q_3$ is a projection operator onto the subspace spanned by all triply excited configurations relative to $|\Phi\rangle$, and the so-called CCSD[T] and CCSD(T) trial ``wave functions'' are given by
\begin{align}
    |\Psi_{\rm CCSD[T]}\rangle &= (1 + T_1 + T_2 + R_0^{(3)}(V_NT_2)_C) |\Phi\rangle, \label{eq:ccsd_t}\\
    |\Psi_{\rm CCSD(T)}\rangle &= |\Psi_{\rm CCSD[T]}\rangle + R_0^{(3)}V_NT_1 |\Phi\rangle. \label{eq:ccsd_T}
\end{align}
Note that the wave functions shown in Eqs. (\ref{eq:ccsd_t}) and (\ref{eq:ccsd_T}) do not provide any 
improvement over the CCSD wave functions, and they are constructed only 
for computing energy corrections through the MMCC formulation. 
Furthermore, the renormalized CCSD[T] and CCSD(T) energy correction can be easily obtained by normalizing $\Delta E_{\rm C
CSD[T]/CCSD(T)}$ with respect to the overlap between CCSD[T]/CCSD(T) trial wave 
functions and CCSD ket wave function, i.e.,
\begin{align}
    \Delta E_{\rm R-CCSD[T]/R-CCSD(T)} &= \frac{\Delta E_{\rm R-CCSD[T]/R-CCSD(T)}}{\langle \Psi_{\rm CCSD[T]/CCSD(T)} | \Psi_{\rm CCSD}\rangle}.
\end{align}
Fig. \ref{fig:HF_results} shows the computed PESs for the HF molecule using the noise-free and 
noisy quantum simulations of the above-mentioned MMCC approaches. As can be seen, in the noise-free simulations, 
the computed MMCC PESs are quite consistent with the classical CCSDTQ results, 
in particular the CCSD and CCSD-l3 results. The (R-)CCSD[T]/ (R-)CCSD(T) results are slightly below the CCSDTQ curve, 
and the deviation becomes slightly larger as approaching long H$-$F distance, which is similar to 
what has been observed in the classical MMCC calculations of 
the HF molecule in the full molecular orbital space reported in Ref. \citenum{mmcc1}. 
In analogy to classical simulations, the renormalized variants prevent the perturbative breakdown of the CCSD[T] and CCSD(T) energies.
Similar curves can also be observed from the noisy simulations except that the computed MMCC PESs 
in the noisy simulations are uniformly shifted upward with respect to the noise-free results by about 0.2 a.u, 
even though in both noise-free and noisy simulations, the equilibrium H$-$F distance exhibited from the computed PESs 
is consistent with the equilibrium H$-$F distance shown in the exact curve. Also, direct measurement of $E_{\rm CCSD-l_3}$ 
become worse as approaching long H$-$F distance, while indirect measurements of $\Delta E_{\rm CCSD[T]/CCSD(T)}$ help 
the computed energies stick close to $E_{\rm CCSD}^{\rm noisy}$, and would be improved if $E_{\rm CCSD}^{\rm noisy}$ gets improved.

\section{Concluding remarks and outlook}
\label{section6}

We have introduced a general quantum algorithm for calculating renormalized CC corrections to energies of 
approximate CC formulations with a non-unitary wave operator. Due to the non-unitary nature of these CC formulations, 
an essential ingredient of our algorithm is to use the occupation information in the SR state to 
represent the non-unitary operator acting on a SR state in terms of equivalent Pauli terms 
without explicitly executing conventional Fermion-to-Pauli transformation. The Pauli terms can further be 
compressed by employing the unitary partitioning approach to obtain a compact unitary basis for a general non-unitary operator. 
%This new approach maps unitary partitioning problem to a graph ``minimum clique cover" problem with the edges of the graph determined by the anti-commuting relationship between Pauli terms, and resorts to the general graph-coloring algorithm for the search of the unitary basis. 
Numerical tests have been conducted for a wide range of systems showing that compact unitary representations 
can be obtained not only for Hamiltonian but also for the left and right CC wave functions, thus exhibiting 
a great potential for the general quantum simulations of non-unitary renormalized CC approaches. Preliminary 
demonstrations of the quantum algorithm are displayed using SIAM and HF systems in both noise-free and noisy simulations. 
The same approach can be straightforwardly extended to approximate the thermal state 
where $e^{-\tau H}$ is the target non-unitary operator. \textcolor{black}{Remarkably,
to approach high accuracy for more strongly correlated systems, one would need to include higher order 
CC amplitudes in the framework to account for higher order many-body effects, whereas our 
unitary partitioning approach will still be very useful in comparison with directly encoding the high-order Taylor expansion of the non-unitary $\exp(T)$ operator.}

In contrast to the algorithms for classical computers, where only non-zero 
low-rank moments can be efficiently evaluated in practical applications, 
quantum computing provides a way of evaluating all non-vanishing moments and enabling more complete forms of 
renormalized CC formulations. Another advantage of the discussed algorithm is its flexibility in handling the situation where an approximate cluster operator is defined adaptively by involving only a selected group of excitations of various ranks. While these adaptive expansions of the approximate cluster operator and corresponding moments pose a significant challenge for classical numerical algorithms, this situation can be easily handled using quantum computers. 
%The MMCC quantum algorithm can also be used to utilize classes of trial wave functions that cannot be easily mapped to classical computers and extend methods of moments to unitary CC formulations. 
The unitary MMCC extension also opens new perspectives for enabling affordable renormalized-type 
corrections for low-circuit-depth UCC-VQE ansatzes and improving approximate VQE energies in a non-iterative manner. 
We will explore these extensions in future papers.

The MMCC formalism is a universal platform for developing energy corrections 
for any ground- and excited-state formalism that uses specific ansatz forms. 
For example, MMCC formalism based on the SR-CC ansatz has been extended to 
excited-state equation-of-motion CC formulations,\cite{creom} 
MR-CC methods,\cite{kowalski2004new,pittner2009method} 
and extended CC approaches,\cite{fan2005non} to mention only a few of its applications. 

\section{ACKNOWLEDGMENTS}
B. P. and K. K. were supported by  the ``Embedding QC into Many-body Frameworks for Strongly Correlated  Molecular and 
Materials Systems'' project, which is funded by the Department of Energy, Office of Science, Office of 
Basic Energy Sciences, the Division of Chemical Sciences, Geosciences, and Biosciences. 
B. P. also acknowledges the support of the Laboratory-Directed Research and Development program from PNNL, 
and the support from the Department of Energy, Office of Science, National Quantum Information Science Research Centers. 

\appendix

\section{Extension of formalism to unitary MMCC ansatz}
\label{MMUCC}

The analysis of the MMCC formalism can be readily extended to employing UCC ansatz. 
To this end we assume ground-state  UCC ansatz 
\begin{equation}
    e^{\sigma^{(A)}} |\Phi\rangle \;,
    \label{mucc1}
\end{equation}
defined by an approximate anti-Hermitian cluster operator $\sigma^{(A)}$
\begin{equation}
    \sigma^{(A)} =T^{(A)}-(T^{(A)})^{\dagger}
    \label{eq9}
\end{equation}
We also assume that the UCC-type trial wave function can be represented in the form
\begin{equation}
    \langle\Psi_T| = \langle\Phi|e^{-\sigma^{(S)}}\;
    \label{eq10}
\end{equation}
where
\begin{equation}
    \sigma^{(S)}=T^{(S)}-(T^{(S)})^{\dagger}
    \label{eq11}
\end{equation}
is parameterized by the cluster operator $T^{(S)}$ that contains higher rank effects than the $T^{(A)}$ operator. 
The unitary MMCC asymmetric energy is  given by the expression
\begin{equation}
    E[\sigma^{(S)}]=
    \frac{\langle\Phi|e^{-\sigma^{(S)}}He^{\sigma^{(A)}}|\Phi\rangle}
    {\langle\Phi|e^{-\sigma^{(S)}}   e^{\sigma^{(A)}}|\Phi\rangle} \;.
    \label{eq12}
\end{equation}
% INCLUDE THIS PART LATER IN THE TEXT
%If vectors $\langle\Phi|e^{-\sigma^{(S)}}$ and $e^{\sigma^{(A)}}|\Phi\rangle$ can be efficiently represented on the quantum registers, then the number of measurements needed to calculate $E(\sigma^{(S)})$ is of the same order as in a single iteration of the VQE formalism. 
%The advantage of the Eq. (\ref{eq12}) is the fact that one does not need explicitly utilize  a specific form of the sufficiency conditions for cluster amplitudes defining the $\sigma^{(A)}$ operator. 
In analogy to the MMCC formulation, functional (\ref{eq12}) can be used to derive a  many-body form of the energy correction. This can be shown by: (1) introducing the resolution of identity 
$e^{\sigma^{(A)}}e^{-\sigma^{(A)}}$
in front of the Hamiltonian $H$ in the numerator of Eq. (\ref{eq12})
\begin{eqnarray}
    E[\sigma^{(S)}]&=&
    \frac{\langle\Phi|e^{-\sigma^{(S)}}
    e^{\sigma^{(A)}}e^{-\sigma^{(A)}}
    He^{\sigma^{(A)}}|\Phi\rangle}
    {\langle\Phi|e^{-\sigma^{(S)}}   e^{\sigma^{(A)}}|\Phi\rangle} \; 
    \nonumber \\
  &=&
  \frac{\langle\Phi|e^{-\sigma^{(S)}}
    e^{\sigma^{(A)}} M^{(A)}_{\rm UCC}|\Phi\rangle}
    {\langle\Phi|e^{-\sigma^{(S)}}   e^{\sigma^{(A)}}|\Phi\rangle} \;,
    \label{eq13} 
\end{eqnarray}
where
\begin{equation}
   M^{(A)}_{\rm UCC}|\Phi\rangle = e^{-\sigma^{(A)}}
    He^{\sigma^{(A)}}|\Phi\rangle \;,
    \label{eq14}
\end{equation}
and (2) introducing the resolution of identity in representation given by  Eq. (\ref{eq5}), i.e.,
\begin{equation}
   E[\sigma^{(S)}] = \frac{\langle\Phi|e^{-\sigma^{(S)}}
    e^{\sigma^{(A)}} (P+Q_A+Q_R)M^{(A)}_{\rm UCC}|\Phi\rangle}
    {\langle\Phi|e^{-\sigma^{(S)}}   e^{\sigma^{(A)}}|\Phi\rangle}  \;.
    \label{eq15}
\end{equation}

The above expression can be rewritten in a form analogous to Eq. (\ref{eq6}), i.e.,
\begin{widetext}
\begin{equation}
E[\sigma^{(S)}] = E^{(A)}_{\rm UCC} +\frac{\langle\Phi|e^{-\sigma^{(S)}}
    e^{\sigma^{(A)}} (Q_A+Q_R)M^{(A)}_{\rm UCC}|\Phi\rangle}
    {\langle\Phi|e^{-\sigma^{(S)}}   e^{\sigma^{(A)}}|\Phi\rangle} \;,
    \label{eq16}
\end{equation}
\end{widetext}
where $E^{(A)}_{\rm UCC}$,
\begin{equation}
    E^{(A)}_{\rm UCC} =\langle\Phi|
    e^{-\sigma^{(A)}} H e^{\sigma^{(A)}} |\Phi\rangle \;.
    \label{eq16}
\end{equation}
is the value of optimized energy in approximate UCC calculations obtained, for example, with the VQE-UCC type simulations on a quantum computer. 
In such a situation,  in contrast to Eq. (\ref{eq6})
where the form of sufficiency conditions for cluster amplitudes can be naturally invoked in the moment expansion, 
in the current form of the UCC formulations, 
all moments corresponding to $Q_A$ projections have  also to be included.  The $Q_A$ projections are not needed in the formula 
(\ref{eq15}) if projective formulations of the UCC sufficiency conditions are employed. 
As discussed in Ref. \citenum{stair2021simulating}, quantum algorithms for the so-called 
projective quantum eigensolver require fewer quantum computational resources than the VQE formulations. 

An interesting alternative to the UCC-type trial wave function in Eq. (\ref{eq10}) is provided by the double unitary CC ansatz
\cite{bauman2019downfolding,downfolding2020t}:
\begin{equation}
    \langle\Psi_T|=\langle\Phi|e^{-\sigma_{\rm int}}
    e^{-\sigma_{\rm ext}}\;,
    \label{eq17}
\end{equation}
where the $\sigma_{\rm int}$ and $\sigma_{\rm ext}$ operators are the so-called internal and external anti-Hermitian operators. 
\textcolor{black}{As discussed in Refs. \citenum{bauman2019downfolding,downfolding2020t,bauman2022coupled,doublec2022}, 
the analysis of the many-body structures of the $\sigma_{\rm int}$ and $\sigma_{\rm ext}$ operators shows that 
they can be approximated in a unitary CC manner
\begin{align}
\sigma_{\rm int} &\approx T_{\rm int} - T_{\rm int}^\dagger, \\
\sigma_{\rm ext} &\approx T_{\rm ext} - T_{\rm ext}^\dagger,
\end{align}
where $T_{\rm int}$ and $T_{\rm ext}$ are SR-CC-type internal (i.e., active) and external cluster operators that satisfy
\begin{align}
T &= T_{\rm int} + T_{\rm ext}. 
\end{align}
The partitioning of the cluster operator $T$ into $T_{\rm int}$ and $T_{\rm ext}$ is based on 
the characterization of sub-systems of a quantum system of interest in terms of the commutative sub-algebras of excitations. 
Specifically,  $T_{\rm int}$ belongs to an excitation sub-algebra that defines the active space 
such that $T_{\rm int}|\Phi\rangle$ produces all Slater determinants in the active space that has the same symmetry as the reference $|\Phi\rangle$.}
%The $\sigma_{\rm int}$ operator is defined through active-space excitations/de-excitations (see Refs. \citenum{bauman2019downfolding,downfolding2020t,bauman2022coupled,doublec2022} for more details).
%
If $\sigma^{(A)}$ is obtained through the optimization of energy functional $E(\sigma^{(A)})$, 
\begin{equation}
    E_{\rm UCC}[\sigma^{(A)}]= \langle\Phi|
    e^{-\sigma^{(A)}} H e^{\sigma^{(A)}} |\Phi\rangle,
    \label{eq18}
\end{equation}
in the active space (i.e. $\sigma^{(A)}$ is defined through cluster amplitudes carrying active spin-orbital indices only) and 
$\sigma^{(A)}\simeq \sigma_{\rm int}$ 
then the moments expansion 
(\ref{eq13}) assumes a simple form:
\begin{equation}
        E[\tilde{\sigma}_{\rm ext}]=
  \langle\Phi|\lbrace e^{-\tilde{\sigma}_{\rm ext}} \rbrace_L
    M^{(A)}_{\rm UCC}|\Phi\rangle
     \;,
    \label{eq19} 
\end{equation}
where general-type anti-Hermitian operator 
$\tilde{\sigma}_{\rm ext}$ is defined as
\begin{equation}
    \tilde{\sigma}_{\rm ext} = -e^{-\sigma^{(A)}} \sigma_{\rm ext}
    e^{\sigma^{(A)}}
    \label{eq20}
\end{equation}
and where the following  identity (see for example Ref. \citenum{szalay1995alternative}) is utilized
\begin{equation}
    \langle\Phi| \lbrace e^{\tilde{\sigma}_{\rm ext}} \rbrace_L = 
    \frac{\langle\Phi|e^{\tilde{\sigma}_{\rm ext}}}
    {\langle\Phi| e^{\tilde{\sigma}_{\rm ext}} |\Phi\rangle} \;.
    \label{eq21}
\end{equation}
In the above expressions, the subscript ``$L$'' designates the linked part of a given operator expression, i.e., part of the expression defined by diagrams that do not contain a disconnected closed part 
\textcolor{black}{(in the traditional terminology of the standard many-body perturbation theory, 
``linked" has for closed diagrams the same meaning as ``connected", while a disconnected open diagram is regarded as ``linked" if it does not include disconnected closed parts, see Refs. \citenum{lindgren2012atomic,Kutzelnigg1984Quantum}).}
It should also be stressed that UCC moments expansion (\ref{eq19}), as long as the approximate  $\sigma_{\rm ext}$ and $\sigma_{\rm int}$/$\sigma^{(A)}$ operators are connected, 
contain connected diagrams only. This is a consequence of the fact that 
the linked part of $e^{-\tilde{\sigma}_{\rm ext}}$ has to be fully contracted with 
the moment operator that includes connected diagrams only. This feature assures 
the proper separation of the energy in the non-interacting sub-system limit 
(or the dissociation limit in the context of  chemical reactions).

\section{Tabulated data for Figs. 1-4}\label{app_table}

\begin{table*}[!htbp]
\resizebox{0.8\linewidth}{!}{%
\begin{tabular}{lllccccccccccccrr}
\hline
\multirow{2}{*}{Molecules}&& \multirow{2}{*}{Basis}&&\multirow{2}{*}{$N_{\rm Qubits}$} && \multicolumn{2}{c}{Hamiltonian} &  & \multicolumn{2}{c}{$\langle \Psi_{\rm CCSD}|$} & &\multicolumn{2}{c}{$| \Psi_{\rm CCSD} \rangle$} & & \multirow{2}{*}{$E_{\rm CCSD}^{\rm classical}$/a.u.} & \multirow{2}{*}{$E_{\rm CCSD}^{\rm noise-free}$/a.u.} \\
\cline{7-8} \cline{10-11} \cline{13-14}
    &&&&&& $N_{\rm Paulis}$ & $N_{\rm Unitaries}$ && $N_{\rm Paulis}$ & $N_{\rm Unitaries}$ && $N_{\rm Paulis}$ & $N_{\rm Unitaries}$ &&&  \\ \hline
~~H$_2$         && STO-3G  &&  4 &&    15 &   11 &&    2 &   2 &&     2 &    2 &&   -1.85141375 &   -1.85141369 \\
~~H$_2$         && 6-31G   &&  8 &&   185 &   50 &&    8 &   4 &&     8 &    4 &&   -1.86581937 &   -1.86581937 \\
~~H$_2$         && cc-pVDZ && 20 &&  2951 &  353 &&   22 &  11 &&    22 &   11 &&   -1.87754224 &   -1.87754224\\
~~H$_4$ (1D)    && STO-3G  &&  8 &&   185 &   50 &&   15 &  11 &&    20 &   12 &&   -3.52635838 &   -3.52635840 \\
~~H$_6$ (1D)    && STO-3G  && 12 &&   919 &  155 &&   60 &  22 &&   191 &   33 &&   -6.06907951 &   -6.06907952 \\
~~H$_8$ (1D)    && STO-3G  && 16 &&  2913 &  383 &&  185 &  36 &&  1691 &  178 &&   -8.85341156 &   -8.85341158 \\
~~H$_{10}$ (1D) && STO-3G  && 20 &&  7151 &  790 &&  442 &  62 && 10572 &  623 &&  -11.82188355 &  -11.82188362 \\
~~H$_4$ (2D)    && STO-3G  &&  8 &&   193 &   56 &&    7 &   5 &&    10 &    6 &&   -3.83614709 &   -3.83614642 \\
~~H$_6$ (2D)    && STO-3G  && 12 &&   919 &  174 &&   98 &  18 &&   345 &   33 &&   -7.26407747 &   -7.26407741 \\
~~H$_8$ (2D)    && STO-3G  && 16 &&  2913 &  376 &&  181 &  33 &&  1677 &  148 &&  -11.36585826 &  -11.36585805 \\
~~H$_{10}$ (2D) && STO-3G  && 20 &&  7147 &  788 &&  876 &  62 && 21126 &  634 &&  -15.69973097 &  -15.69973098 \\
~~H$_4$ (3D)    && STO-3G  &&  8 &&   133 &   37 &&   27 &   5 &&    36 &    6 &&   -3.62075876 &   -3.62075905 \\
~~H$_6$ (3D)    && STO-3G  && 12 &&   503 &   93 &&   60 &  11 &&   191 &   20 &&   -7.59028941 &   -7.59028942 \\
~~H$_8$ (3D)    && STO-3G  && 16 &&  3609 &  378 &&  301 &  46 &&  3187 &  227 &&  -12.34981407 &  -12.34981386 \\
~~H$_{10}$ (3D) && STO-3G  && 20 &&  5159 &  509 &&  674 &  41 && 17146 &  221 &&  -17.32138157 &  -17.32138157 \\
~~LiH           && STO-3G  && 12 &&   631 &  104 &&   35 &   9 &&    69 &    9 &&   -8.87777296 &   -8.87777295 \\
~~HF            && STO-3G  && 12 &&   631 &  104 &&   18 &   7 &&    18 &    7 && -103.79138904 & -103.79138904 \\
~~BeH$_2$       && STO-3G  && 14 &&   666 &  118 &&   39 &  13 &&   143 &   18 &&  -18.98592175 &  -18.98592184 \\
~~H$_2$O        && STO-3G  && 14 &&  1086 &  180 &&   49 &  11 &&   133 &   16 &&  -84.20199538 &  -84.20199546 \\
~~H$_4$ (1D)    && 6-31G   && 16 &&  2913 &  388 &&  103 &  18 &&   400 &   35 &&   -3.63695844 & \\
~~H$_4$ (1D)    && cc-pVDZ && 40 && 53289 & 1870 &&  415 &  40 &&  7780 &   83 &&   -3.65296044 & \\
~~H$_6$ (1D)    && 6-31G   && 24 && 14905 & 1376 &&  504 &  45 && 11608 &  159 &&   -6.23248485 & \\
~~BH$_3$        && STO-3G  && 16 &&  1973 &  306 &&  117 &  19 &&   829 &   33 &&  -33.56240621 & \\
~~NH$_3$        && STO-3G  && 16 &&  3041 &  435 &&  164 &  20 &&  1228 &   37 &&  -67.47299891 & \\
~~CH$_4$        && STO-3G  && 18 &&  6892 &  553 &&  429 &  43 &&  7369 &  110 &&  -53.27748972 & \\
~~O$_2$         && STO-3G  && 20 &&  2239 &  270 &&   55 &  14 &&   231 &   29 && -175.74286915 & \\
~~F$_2$         && STO-3G  && 20 &&  2951 &  351 &&   22 &   9 &&    22 &    9 && -226.40827516 & \\
~~N$_2$         && STO-3G  && 20 &&  2951 &  353 &&  136 &  21 &&  1485 &   50 && -131.27286827 & \\
~~HCl           && STO-3G  && 20 &&  5851 &  536 &&   48 &   7 &&    48 &    7 && -462.21179568 & \\
~~CO            && STO-3G  && 20 &&  5851 &  539 &&  234 &  33 &&  3521 &  380 && -133.86962585 & \\
~~NO$^+$        && STO-3G  && 20 &&  5851 &  539 &&  234 &  33 &&  3525 &  386 && -153.14592464 & \\
~~NaH           && STO-3G  && 20 &&  5851 &  554 &&  351 &  45 &&  7487 &   76 && -163.44057301 & \\
~~MgH$_2$       && STO-3G  && 22 &&  4582 &  481 &&  249 &  35 &&  6363 &   55 && -205.75425895 & \\
~~H$_2$S        && STO-3G  && 22 &&  6246 &  711 &&  141 &  13 &&   829 &   18 && -407.30831861 & \\
~~AlH$_3$       && STO-3G  && 24 &&  8753 &  750 &&  415 &  37 && 13359 &   46 && -254.43679545 & \\
~~PH$_3$        && STO-3G  && 24 && 15233 & 1153 &&  566 &  29 && 15468 &   42 && -356.22480747 & \\
~~SiH$_4$       && STO-3G  && 26 && 33648 & 1939 &&  417 &  80 && 83345 &  125 && -309.32192908 & \\
~~CO$_2$        && STO-3G  && 30 && 16170 & 1359 &&  599 &  68 && 41077 & 2818 && -243.54345847 & \\
~~NO$_2^+$      && STO-3G  && 30 && 20570 & 1739 &&  735 & 186 && 47863 & 3114 && -266.27028079 & \\
~~N$_2$O        && STO-3G  && 30 && 31634 & 2108 && 1185 & 299 && 88645 & 5479 && -334.97878855 & \\
~~Cl$_2$        && STO-3G  && 36 && 34591 & 2468 &&   74 &  10 &&    74 &   11 && -986.07096594 & \\ \hline
\end{tabular}
}
\caption{Tabulated data for Figs. 1 and  2 in the main text.}
\label{tab:comparison1}
\end{table*}

\begin{table*}[!htbp]
\resizebox{\linewidth}{!}{%
\begin{tabular}{ccccccccccc}
\hline
\multirow{2}{*}{$\log_{10}(U/V)$} && \multicolumn{3}{c}{Noise-free} && \multicolumn{3}{c}{Noisy ($FakeToronto$)} && \multirow{2}{*}{$E_{\rm Exact}$} \\ \cline{3-5} \cline{7-9}
&& $E_{\rm CCSD}$ & $E_{\rm CCSD-\Lambda_3}$ & $E_{\rm CCSDT}$ && $E_{\rm CCSD}$ & $E_{\rm CCSD-\Lambda_3}$ & $E_{\rm CCSDT}$ && \\  \hline 
0.0 &&  -5.15881517  &  -5.15891923 &  -5.15891987 &&  -4.94555961$\pm$0.00442462 & -4.97146118$\pm$0.00391061 & -4.94104491$\pm$0.00464566 && -5.15891987 \\
0.3 &&  -5.43674925  &  -5.43719446 &  -5.43719792 &&  -5.21784849$\pm$0.00396464 & -5.25543809$\pm$0.00498839 & -5.21816191$\pm$0.00569655 && -5.43719790 \\
0.6 &&  -6.04647846  &  -6.04849319 &  -6.04851725 &&  -5.82402531$\pm$0.00346324 & -5.87346886$\pm$0.00328429 & -5.82211880$\pm$0.00557078 && -6.04851697 \\
0.9 &&  -7.45178382  &  -7.46307880 &  -7.46340952 &&  -7.19474708$\pm$0.00583541 & -7.27061309$\pm$0.00498225 & -7.20342213$\pm$0.00527000 && -7.46340418 \\
1.2 && -10.78419849  & -10.79953788 & -10.79986846 && -10.44935554$\pm$0.00946826 &-10.47284980$\pm$0.01162278 &-10.46315207$\pm$0.01162418 && -10.79981136 \\
1.5 && -18.20429450  & -18.24533312 & -18.24730462 && -17.66700160$\pm$0.02755633 &-17.45479343$\pm$0.02034044 &-17.70924716$\pm$0.02468933 && -18.24713845 \\ \hline
\end{tabular}}
\caption{Tabulated direct measurement data for Fig. 3 in the main text. 
The each noisy data point is an average of 10 experiment with 10,000 shots for each experiment.}
\label{tab:comparison2}
\end{table*}

\begin{table*}[!htbp]
\resizebox{\linewidth}{!}{%
\begin{tabular}{ccccccccc}
\hline
\multirow{2}{*}{$\log_{10}(U/V)$} && \multicolumn{3}{c}{Noise-free} && \multicolumn{3}{c}{Noisy ($FakeToronto$)} \\ \cline{3-5} \cline{7-9}
&& $\langle \Phi |\Lambda_3 e^{-T}|H|e^T|\Phi\rangle$ & $\langle \Phi| \Lambda_3 e^{-T}|\cdot|e^T|\Phi\rangle$ & $E_{\rm CCSD-\Lambda_3}$ && $\langle \Phi |\Lambda_3 e^{-T}|H|e^T|\Phi\rangle$ & $\langle \Phi |\Lambda_3 e^{-T}|\cdot|e^T|\Phi\rangle$ & $E_{\rm CCSD-\Lambda_3}$ \\  \hline 
0.0 &&  -0.03111908  &  0.00601192 &  -5.15891923 &&  -0.02973900$\pm$3.6318e-05 & 0.00601192$\pm$1.9778e-18 & -5.157547408 \\
0.3 &&  -0.03842588  &  0.00698534 &  -5.43719446 &&  -0.03681447$\pm$3.6139e-05 & 0.00698534$\pm$1.8602e-18 & -5.435594246 \\
0.6 &&  -0.05366766  &  0.00853980 &  -6.04849319 &&  -0.05164359$\pm$4.0003e-05 & 0.00853980$\pm$1.7347e-18 & -6.046486267 \\
0.9 &&  -0.07763773  &  0.00888945 &  -7.46307880 &&  -0.07487089$\pm$5.9288e-05 & 0.00888945$\pm$4.4565e-18 & -7.460336422 \\
1.2 &&  -0.03906846  &  0.00219721 & -10.79953788 &&  -0.03760697$\pm$6.8678e-05 & 0.00219721$\pm$7.5115e-19 & -10.79807981 \\
1.5 &&   0.22012806  & -0.01431416 & -18.24533312 &&   0.21884322$\pm$0.00026832 &-0.01431416$\pm$2.6874e-18 & -18.24663656 \\ \hline
\end{tabular}}
\caption{Tabulated indirect measurement data for obtaining noisy $E_{\rm CCSD-\Lambda_3}$, shown in Fig. 3 in the 
main text, 
using $E_{\rm CCSD-\Lambda_3} = \frac{\langle \Psi_T|H|e^{T}|\Phi\rangle}{\langle\Psi_T|e^{T}|\Phi\rangle} = \frac{E_{\rm CCSD}+\langle \Phi |\Lambda_3 e^{-T}|H|e^T|\Phi\rangle}{1 + \langle \Phi |\Lambda_3 e^{-T}|\cdot|e^{T}|\Phi\rangle}$, 
where $E_{\rm CCSD}$ is obtained from classical computing.}
\label{tab:comparison3}
\end{table*}

\begin{table*}[!htbp]
\resizebox{\linewidth}{!}{%
\begin{tabular}{cccccccccc}
\hline
Dist./$\AA$ && $E_{\rm HF}^{\rm classical}$/a.u. & $E_{\rm CCSDTQ}^{\rm classical}$/a.u. & $E_{\rm CCSD}^{\rm noise-free}$/a.u. & $E_{\rm CCSD-l_3}^{\rm noise-free}$/a.u. & $E_{\rm R-CCSD(T)}^{\rm noise-free}$/a.u. & $E_{\rm R-CCSD[T]}^{\rm noise-free}$/a.u. & $E_{\rm CCSD(T)}^{\rm noise-free}$/a.u. & $E_{\rm CCSD[T]}^{\rm noise-free}$/a.u. \\ \hline
0.6 && -99.72732749 & -99.74924268 & -99.74898034 & -99.74915658 & -99.74915496 & -99.74917596 & -99.74915627 & -99.74917744 \\
0.7 && -99.92096796 & -99.95795651 & -99.95745811 & -99.95789116 & -99.95789623 & -99.95797043 & -99.95790176 & -99.95797690 \\
0.8 && -99.99948359 & -100.0394432 & -100.0388268 & -100.0393763 & -100.0393888 & -100.0395000 & -100.0393966 & -100.0395093 \\
0.9 && -100.0215114 & -100.0658488 & -100.0650373 & -100.0657766 & -100.0658018 & -100.0659798 & -100.0658142 & -100.0659950 \\
1.0 && -100.0157879 & -100.0576983 & -100.0572387 & -100.0574741 & -100.0574754 & -100.0575473 & -100.0574800 & -100.0575533 \\
1.5 && -99.89198356 & -99.96622946 & -99.96502878 & -99.96553766 & -99.96557304 & -99.96590984 & -99.96562518 & -99.96599418 \\
2.0 && -99.78357371 & -99.90110675 & -99.89933579 & -99.90038410 & -99.90071607 & -99.90174961 & -99.90123664 & -99.90265966 \\
2.5 && -99.71145146 & -99.88559991 & -99.88407156 & -99.88574201 & -99.88688445 & -99.88858604 & -99.88938554 & -99.89260031 \\
3.0 && -99.66583950 & -99.88181130 & -99.88094675 & -99.88229220 & -99.88486955 & -99.88707153 & -99.89001862 & -99.89511433 \\
3.5 && -99.63767250 & -99.88037159 & -99.88005052 & -99.88062028 & -99.88446597 & -99.88695172 & -99.89119194 & -99.89747101 \\
4.0 && -99.62040722 & -99.88030227 & -99.88021870 & -99.88037685 & -99.88473591 & -99.88728617 & -99.89190712 & -99.89851428 \\
4.5 && -99.60941181 & -99.88063715 & -99.88063127 & -99.88066272 & -99.88506557 & -99.88757507 & -99.89210310 & -99.89860372 \\
5.0 && -99.60179863 & -99.88097821 & -99.88098817 & -99.88098807 & -99.88949143 & -99.89433798 & -99.90282578 & -99.91530392 \\ \hline
\end{tabular}}
\caption{Tabulated noise-free energy points shown in Fig. 4 in the main text.}
\label{tab:comparison4}
\end{table*}

\begin{table*}[!htbp]
\resizebox{\linewidth}{!}{%
\begin{tabular}{cccccccc}
\hline
Dist./$\AA$ && $E_{\rm HF}^{\rm classical}$/a.u. & $E_{\rm CCSDTQ}^{\rm classical}$/a.u. & $E_{\rm CCSD}^{\rm noisy}$/a.u. & $E_{\rm CCSD-l_3}^{\rm noisy}$/a.u. & $\Delta E_{\rm CCSD[T]}^{\rm noisy}$/a.u. & $\Delta E_{\rm CCSD(T)}^{\rm noisy}$/a.u. \\ \hline
0.6 && -99.72732749	& -99.74924268 & -99.54199096$\pm$0.00625921 & -99.54070258$\pm$0.01447560 & -0.00019709 & -0.00017593 \\
0.7 && -99.92096796	& -99.95795651 & -99.75146185$\pm$0.01434792 & -99.75412702$\pm$0.00924627 & -0.00051878 & -0.00044365 \\
0.8 && -99.99948359	& -100.0394432 & -99.84311069$\pm$0.01942037 & -99.82853624$\pm$0.01419339 & -0.00068253 & -0.00056983 \\
0.9 && -100.0215114	& -100.0658488 & -99.86724381$\pm$0.01357963 & -99.87431445$\pm$0.01215604 & -0.00095767 & -0.00077686 \\
1.0 && -100.0157879	& -100.0576983 & -99.86636387$\pm$0.01430875 & -99.85613148$\pm$0.00724880 & -0.00031463 & -0.00024129 \\
1.5 && -99.89198356	& -99.96622946 & -99.77855024$\pm$0.01298849 & -99.78768765$\pm$0.00613690 & -0.00096540 & -0.00059640 \\
2.0 && -99.78357371	& -99.90110675 & -99.69697254$\pm$0.01959051 & -99.70498621$\pm$0.01218729 & -0.00332386 & -0.00190085 \\
2.5 && -99.71145146	& -99.88559991 & -99.68477006$\pm$0.01530131 & -99.66957651$\pm$0.01029929 & -0.00852874 & -0.00531397 \\
3.0 && -99.6658395	& -99.88181130 & -99.68302408$\pm$0.01587450 & -99.63453273$\pm$0.01182176 & -0.01416758 & -0.00907186 \\
3.5 && -99.6376725	& -99.88037159 & -99.67854856$\pm$0.00874610 & -99.6004408$\pm$0.015724201 & -0.01742048 & -0.01114142 \\
4.0 && -99.62040722	& -99.88030227 & -99.67302256$\pm$0.01819842 & -99.60126092$\pm$0.01668604 & -0.01829557 & -0.01168842 \\
4.5 && -99.60941181	& -99.88063715 & -99.67514965$\pm$0.01215763 & -99.59727933$\pm$0.01180143 & -0.01797244 & -0.01147182 \\
5.0 && -99.60179863	& -99.88097821 & -99.67912429$\pm$0.01710346 & -99.52965464$\pm$0.00916324 & -0.03431575 & -0.02183761 \\ \hline
\end{tabular}}
\caption{Tabulated noisy energy points shown in Fig. 4 in the main text. 
The each noisy data point is an average of 10 experiment with 10,000 shots for each experiment. 
The computed standard deviations for $\Delta E_{\rm CCSD[T]}^{\rm noisy}$ and $\Delta E_{\rm CCSD(T)}^{\rm noisy}$ are too small 
(i.e., $\sim10^{18}$) to be included.}
\label{tab:comparison5}
\end{table*}

\bibliography{ref}
\end{document}